\documentclass[12pt]{article}
\usepackage{graphicx}
\usepackage{amsmath}
\usepackage{float}
\usepackage{color}

\def\RR{{\rm I\kern-.17em R}}

\begin{document}
\title{GEODESICS DYNAMICS IN THE LINET-TIAN SPACETIME WITH $\Lambda<0$}
\author{Irene Brito$^1$\thanks{e-mail: ireneb@math.uminho.pt}, M. F. A. Da Silva$^2$\thanks{e-mail: mfasnic@gmail.com}, Filipe C. Mena$^{1,2}$\thanks{e-mail: fmena@math.uminho.pt},\\
and N. O. Santos$^{3,4}$\thanks{e-mail: n.o.santos@qmul.ac.uk}\\
{\small $^1$Centro de Matem\'atica, Universidade do Minho, 4710-057 Braga, Portugal.}\\
{\small $^2$Departamento de F\'{i}sica Te\'orica, Instituto de F\'isica,}\\
{\small Universidade do Estado do Rio de Janeiro, Rua S\~ao Francisco Xavier 524,}\\
{\small Maracan\~a, 20550-900, Rio de Janeiro, Brazil.}\\
{\small $^3$School of Mathematical Sciences, Queen Mary,}
{\small University of London, London E1 4NS, UK.}\\
{\small $^4$Observatoire de Paris, Universit\'e Pierre et Marie Curie,} \\
{\small LERMA(ERGA) CNRS - UMR 8112, 94200 Ivry sur Seine, France.}}
\maketitle

\begin{abstract}
We investigate the geodesics' kinematics and dynamics in the
Linet-Tian metric with $\Lambda<0$ and compare with the results for
the Levi-Civita metric, when $\Lambda=0$. This is used to derive new stability results about the geodesics' dynamics in static vacuum
cylindrically symmetric spacetimes with respect to the introduction
of $\Lambda<0$.

In particular, we find that increasing $|\Lambda|$
always increases the minimum  and maximum radial distances to the axis of any spatially confined planar null geodesic.
Furthermore, we show that, in some cases, the inclusion of any $\Lambda<0$ breaks the geodesics' orbit confinement of the $\Lambda=0$ metric, for both planar and non-planar null geodesics, which are therefore unstable.

Using the full system of geodesics' equations, we provide numerical examples which illustrate our results.
\end{abstract}

\section{Introduction}
The static vacuum spacetime that describes the exterior to an infinite cylinder of matter is the Levi-Civita (LC) spacetime \cite{LC}.
In its general form, it contains two independent parameters \cite{Bonnor,Bonnor1,Wang}, one is the Newtonian mass per unit length,
usually  denoted by $\sigma$, and the other is associated an angle defect. The generalisation of this spacetime to include a non-zero cosmological constant $\Lambda$,
which can be either positive or negative, has been obtained by Linet \cite{L} and Tian \cite{T}, and it has been shown
by Da Silva et al. \cite{Silva-etal} and Griffiths and Podolsk\'{y} \cite{GP} that some properties of the LC spacetime are dramatically modified.

Cylindrically symmetric spacetimes have a
wide range of importance to study  several physical systems (see e.g. \cite{Brito} and references there in). In particular,
the Linet-Tian (LT) spacetime has been used to describe e.g. cosmic strings \cite{T,Bezerra, Bhattacharya} and was found to be the unique exterior to some static spacetimes \cite{Brito, Brito2,GP,Zofka-Bicak}.

The purpose of this paper is to delve further in understanding the properties of the LT spacetime by restricting to the case $\Lambda<0$.
In order to do this, we endeavour more deeply into the study of the kinematics and dynamics of its geodesics.

The study of the stability of geodesics around axially symmetric spacetimes in General Relativity is an old problem (see e.g. \cite{Bardeen}) and it is specially relevant e.g. in the study of the motion of photons and other test particles around astrophysical objects. Two sets of problems connected to this issue, which are still object of research, are the analysis of chaotic motion of geodesics in non-spherical metrics (see e.g. \cite{Gueron}) and the question of the stability of closed timelike curves (see \cite{Rosa,Gron}).

However, little has been done about the stability of geodesics with respect to the introduction of a $\Lambda$ term in the Einstein field equations.
Banerjee et al. \cite{Banerjee} were the first to consider
the study of geodesics in LT spacetimes by investigating the dynamics of planar geodesics in terms of the constant $\sigma$.
In particular, they derived conditions under which null and timelike geodesics are confined or may escape to infinity. In some cases, they have also compared their results with the LC case studied in \cite{Banerjee0,Silva1, Bonnor3, Gautreau}.

Here, we extend the results of \cite{Banerjee}, giving a clearer view of the parameters involved by defining an appropriate effective potential and by analysing the dynamics of not only planar, but also non-planar geodesics. Also, differently from \cite{Banerjee}, in some occasions, we use a linear perturbative approach which allows us, in a  mathematically more precise way, to study the effects on the orbits of the introduction of an arbitrarily small $\Lambda$.

Our results also generalize some results obtained in \cite{Herrera2} for the Lewis metric in the limit of static spacetimes. Previous results for the LC  spacetime are recovered here in the limit $\Lambda\rightarrow 0$ and we discuss the differences between the geodesics' dynamics in the LT and LC spacetimes. In particular, we investigate how the introduction of $\Lambda$ interferes with the stability of the geodesics in the LC spacetime.

The plan of the paper is the following: In Section 2, we recall the geodesics' system for the LT metric. Section 3 deals with circular geodesics and, in particular, we look at the stability of the geodesics' proper radius with respect to the introduction of a linear perturbation in $\Lambda$. In Section 4, we study geodesics along the symmetry axis direction and we emphasise the differences between the LT and LC cases. In Section 5, we investigate in detail the dynamics of geodesics along the radial direction. We define a potential which we use in order to study, separately, planar and non-planar geodesics in LT. In each case, we split our analysis into null and non-null geodesics, we compare our results with the $\Lambda=0$ case and we illustrate our results by plotting numerical simulations of the geodesics dynamics. We finish the paper with a brief conclusion. Throughout, we use units such that $8\pi G$=c=1.

\section{Geodesics in the LT spacetime}

The LT metric can be expressed as \cite{L,T}
\begin{equation}
ds^2=-fdt^2+d\rho^2+gdz^2+ld\phi^2, \label{1}
\end{equation}
with
\begin{eqnarray}
f=Q^{2/3}P^{-2(1-8\sigma+4\sigma^2)/3\Sigma}, \label{2}\\
g=Q^{2/3}P^{-2(1+4\sigma-8\sigma^2)/3\Sigma}, \label{3}\\
l=c^2Q^{2/3}P^{4(1-2\sigma-2\sigma^2)/3\Sigma}, \label{four}
\end{eqnarray}
where $t$, $\rho$, $z$ and $\phi$ are the usual cylindrical coordinates, $\Sigma=1-2\sigma+4\sigma^2$, the constant $\sigma$ is related, but not equal, to the mass per unit length, the constant $c>0$ is related to the angle defect \cite{Bonnor,Bonnor1,Wang} and, for $\Lambda<0$,
\begin{equation}
P=\frac{2}{\sqrt{3|\Lambda|}}\tanh R, \;\; Q=\frac{1}{\sqrt{3|\Lambda|}}\sinh(2R), \;\;
R=\frac{\sqrt{3|\Lambda|}}{2}\rho\,. \label{5}
\end{equation}
In the limit $\Lambda\rightarrow 0$, the metric reduces to the LC metric for which $P=Q=\rho$.

For $1/2\leq\sigma<\infty$, it is known that the value of $\sigma$ makes the axial and angular coordinates switch meaning  \cite{Herrera3}. For $1/2<\sigma<\infty$, the spacetime description appears to be similar to the $0\leq\sigma\leq 1/2$ case by redefining $\sigma$. For this reason, we assume the range of $\sigma$ to be $0\leq\sigma\leq 1/2$.

The LT spacetime with $\sigma=0$ and $\Lambda\neq 0$ has some similar characteristics to the de Sitter and anti-de Sitter spacetimes, since these spacetimes have a zero energy momentum tensor.
However, for $\Lambda\neq 0$, the LT spacetime does not reduce to the de Sitter or anti-de Sitter spacetimes as $\sigma\rightarrow 0$,
since both spacetimes are not compatible with static cylindrical symmetry. In fact, if the de Sitter and anti-de Sitter spacetimes
are expressed in cylindrical coordinates, then the metrics are explicitly time dependent \cite{Bonnor2},
and are of Petrov type D as shown in \cite{Silva-etal}.
Bonnor \cite{Bonnor2} called the LT spacetime with $\Lambda<0$ and $\sigma=0$ {\em non-uniform anti-de Sitter spacetime}.

In what follows, we leave the parameters $0\le\sigma\le 1/2$ and $c>0$ as general as possible. However, some of our calculations will not hold in the cases $\sigma=0,1/4,1/2$ and these will be treated separately whenever necessary. In fact, for $\sigma=0,1/4,1/2$ the spacetime is of Petrov type D and has some interesting distinctive properties from the remaining cases, see e.g. \cite{Silva-etal, GP}.
The case  $\sigma=1/2$, in particular, upon a coordinate transformation to planar symmetry gives rise to the so called {\em black membranes} \cite{Silva-etal}. In this paper, however, we will focus on cylindrically symmetric geometries in which case, as in the LC spacetime, there are no trapped cylinders as we will show next.

We recall that a trapped surface is a 2-dimensional imbedded spatial surface such that its
causal future is (at least initially) contained within regions of
decreasing area. Given a surface $S$, one can establish whether it is
trapped or not by studying the traces $\theta^\pm$ of the null second fundamental forms on $S$ defined below
\cite{Penrose,Senovilla}. In particular, $S$ is a trapped surface if
$\theta=2\theta^+\theta^->0$, marginally trapped if $\theta=0$ and
untrapped if $\theta<0$. So, consider 2-surfaces $S$
spanned by the vectors $\vec e_1=\partial_\phi$ and $\vec e_2=\partial_z$. One can then define
future-directed null vectors orthogonal to $S$ as
\begin{equation}
\vec k^\pm=\frac{\sqrt{2}}{2}\left(\frac{1}{\sqrt{f}}\,\partial_t\pm\partial_\rho\right),
\end{equation}
compute the second fundamental forms $\theta_{AB}^\pm=-k_\mu^\pm e_A^\nu\nabla_\nu e_B^\mu$, on $S$, with $A,B=1,2$, and their traces by using (\ref{1}),
\begin{equation}
\theta^\pm=\mp\frac{\sqrt{2}}{6}\frac{1}{PQ\Sigma}\left[2Q'P\Sigma+QP'(4\sigma^2-8\sigma+1)\right]
\end{equation}
giving
\begin{equation}
\theta^+\theta^-< 0, ~~{\mbox {for}}~~ \rho\in ~]0,\infty[,
\end{equation}
 which indicates that there are no trapped cylinders, in this case.
However, as was pointed out in \cite{Banerjee}, there exist families of trapped null planar geodesics and we will explore those aspects in more detail ahead also generalising some of their results to the non-planar case.

From (\ref{1}), we can obtain the geodesics equations (see also \cite{Herrera2,Silva1})
\begin{eqnarray}
{\ddot t}+\frac{f^{\star}}{f}{\dot t}{\dot \rho}=0, \label{7}\\
2{\ddot\rho}+f^{\star}{\dot t}^2-g^{\star}{\dot z}^2-l^{\star}{\dot\phi}^2=0, \label{8}\\
{\ddot z}+\frac{g^{\star}}{g}{\dot\rho}{\dot z}=0, \label{9}\\
{\ddot\phi}+\frac{l^{\star}}{l}{\dot\rho}{\dot\phi}=0, \label{10}
\end{eqnarray}
where the dot and star stand, respectively, for differentiation with
respect to an affine parameter $\lambda\ge 0$ and the coordinate
$\rho\ge 0$. After integrating (\ref{7})-(\ref{10}), we obtain
\begin{eqnarray}
{\dot t}=\frac{E}{f}, \label{11}\\
{\dot\rho}^2=\frac{E^2}{f}-\epsilon-\frac{P_z^2}{g}-\frac{L_z^2}{l}, \label{12}\\
{\dot z}=\frac{P_z}{g}, \label{13}\\
{\dot\phi}=\frac{L_z}{l}, \label{14}
\end{eqnarray}
 where $\epsilon=0, 1, -1$ for null, timelike and spacelike geodesics, respectively, and the constants $E$, $P_z$ and $L_z$
represent, respectively, the total energy of the test particle, its
momentum along the $z$ axis and its angular momentum about the $z$
axis, which are all assumed to be finite for $\rho\in  ]0, \infty[$.

\section{Circular planar geodesics ($\dot \rho=\dot z = 0$)}

Circular geodesics for $\Lambda<0$ were investigated in \cite{Banerjee}. Interestingly, they found that $\sigma$ determines whether the geodesics are timelike, null or spacelike,  independently of their radial distance to the axis \cite{Banerjee}. A similar property was already known for the LC spacetime for which, in the corresponding cases, $\sigma$ is lower, equal or greater than $1/4$.

In this section, we focus on the changes introduced in the geodesics' dynamics with the inclusion of $\Lambda$ by studying, in detail, the tangential velocity and acceleration as well as the geodesics' proper radius.

\subsection{Tangential velocity and acceleration}

 We restrict our study to circular geodesics in the plane perpendicular to $z$, in which case ${\dot\rho}={\dot z}=0$,
and it is easy to integrate (\ref{11}) and (\ref{14}) to get
simply $t=(E/f)\lambda$ and $\phi=(L_z/l)\lambda$ or
$\phi=\left(fL_z)/(lE\right)t$.

From
(\ref{8}), we get
\begin{equation}
\omega^2=\left(\frac{\dot\phi}{\dot t}\right)^2=\frac{f^{\star}}{l^{\star}}, \label{16}
\end{equation}
where $\omega$ defines the angular velocity of the particle along a geodesic around the $z$ axis, whereas its tangential velocity $W$  is given by \cite{Silva1,Herrera2}
\begin{equation}
W^2=\frac{l}{f}\,\omega^2. \label{17}
\end{equation}
Substituting (\ref{5}) and (\ref{16}) into (\ref{17}) we obtain
\begin{equation}
W^2=\frac{2\Sigma\sinh^2R+6\sigma}{2\Sigma\sinh^2R+3(1-2\sigma)}=c^2P^{2(1-4\sigma)/\Sigma}\omega^2, \label{18}
\end{equation}
 having the proper Newtonian limit $W=\rho\omega$, with $c=1$.
When $\Lambda=0$, i.e. in the LC spacetime, (\ref{18}) becomes
\begin{equation}
W^2_{LC}=\frac{2\sigma}{1-2\sigma}, \label{19}
\end{equation}
which does not depend on $\rho$ meaning that, for every $0\le \sigma< 1/2$, the circular tangential velocity is fixed in the range $ 0<\rho<\infty$.

If $|\Lambda|\ll1$, then up to first order in $|\Lambda|$, we obtain from (\ref{18})
\begin{equation}
W^2\approx W_{LC}^2+|\Lambda|\frac{\Sigma}{2}\frac{1-4\sigma}{(1-2\sigma)^2}\rho^2, \label{19a} ~~\text{for}~~\sigma\ne \frac{1}{2},
\end{equation}
showing explicitly how, for $0\le \sigma<1/4$, at linear order, $|\Lambda|$ increases the corresponding tangential velocity for the LC circular geodesics, while by increasing $\rho$, increases the tangential velocity too. However, $\sigma>1/4$ has the opposite effect.

When $\sigma=1/4$, then from (\ref{18}), $|\Lambda|$ has no influence upon the corresponding LC tangential velocity which becomes $W_{LC}=1$.

Furthermore, near the axis, $\rho\ll 1$, $W$ reduces to $W_{LC}$ since the LT metric reduces to the LC metric (see also \cite{GP}).
From (\ref{18}), as $\rho\rightarrow\infty$, we have that $W^2\rightarrow 1$ for any $0<\sigma<1/2$.

By differentiating (\ref{18}) with respect to $\rho$ we obtain
\begin{equation}
\label{star}
W^{2\star}=\frac{3(1-4\sigma)\Sigma\sqrt{3|\Lambda|}\sinh(2R)}{[\Sigma \cosh(2R)+2(1-2\sigma-2\sigma^2)]^2},
\end{equation}
showing that, for $0<\sigma<1/4$,  we get $W^{2\star}>0$, i.e. the tangential velocity is an increasing function of $\rho$ while, for $\sigma>1/4$, $W^{2\star}<0$ and the tangential velocity is a decreasing function of $\rho$.

For the special cases $\sigma=0,1/4,1/2$, whose physical relevance is highlighted e.g. in \cite{GP}, the following results emerge: For $\sigma=0$, we have from (\ref{18}) and (\ref{star}),
\begin{equation}
W^2_0=\frac{\cosh(2R)-1}{\cosh(2R)+2}<1, ~~~~W^{2\star}_0=\frac{3\sqrt{3|\Lambda|}\sinh {(2R)}}{2\sinh^2{R}+3}>0, \label{21}
\end{equation}
showing that if $\rho\rightarrow 0$, then $W_0^2\rightarrow 0$ and $W^{2\star}_0\rightarrow 0$, while if $\rho\rightarrow\infty$, then $W_0^2\rightarrow 1$ and $W^{2\star}_0\rightarrow 1$. For $\sigma=1/4$, we have
\begin{equation}
W^2_{1/4}=1,~~~~W^{2\star}_{1/4}=0, \label{22}
\end{equation}
which means that, independently of the $\rho$ distance and the $|\Lambda|$ value, the circular geodesics are null in this case, like in the LC spacetime, as observed in (\ref{19a}).
For $\sigma=1/2$, we have
\begin{equation}
W^2_{1/2}=\frac{\cosh(2R)+2}{\cosh(2R)-1}>1, ~~~~W^{2\star}_{1/2}=-\frac{3\sqrt{3|\Lambda|}\coth{R}}{2\sinh^2{R}}<0, \label{23}
\end{equation}
where the tangential velocity for $\rho\rightarrow 0$ becomes $W_{1/2}^2\rightarrow \infty$ and $W^{2\star}_{1/2}\rightarrow -\infty$, while if $\rho\rightarrow\infty$, then $W_{1/2}^2\rightarrow 1$ and $W^{2\star}_{1/2}\rightarrow -\infty$.

\subsection{Proper radius}

\begin{figure}[H]
\begin{minipage}{8cm}
\begin{tabular}{c}
\includegraphics[width=8cm]{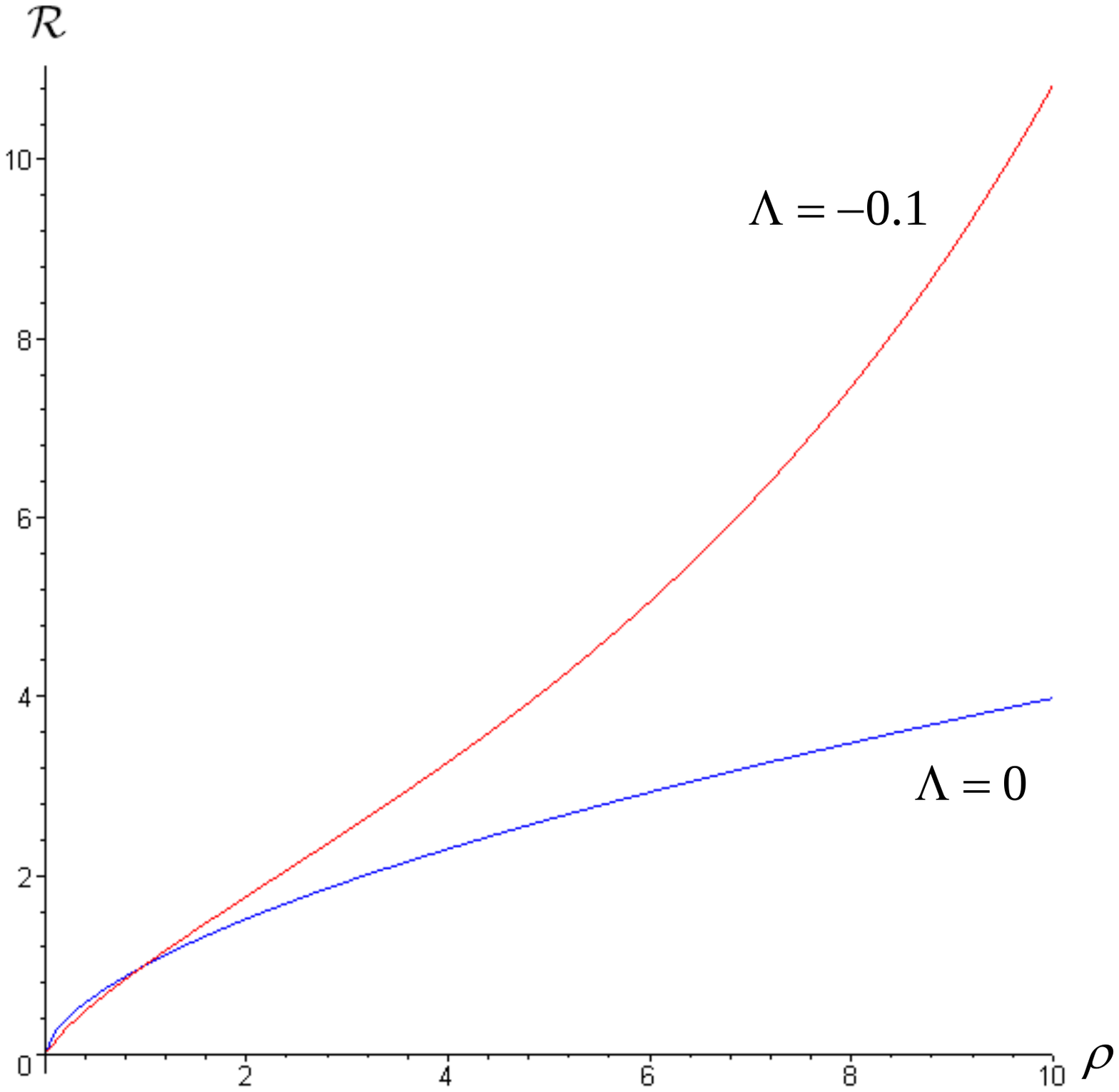} \\
\end{tabular}
\end{minipage}
\begin{minipage}{8cm}
\begin{tabular}{c}
\includegraphics[width=8cm]{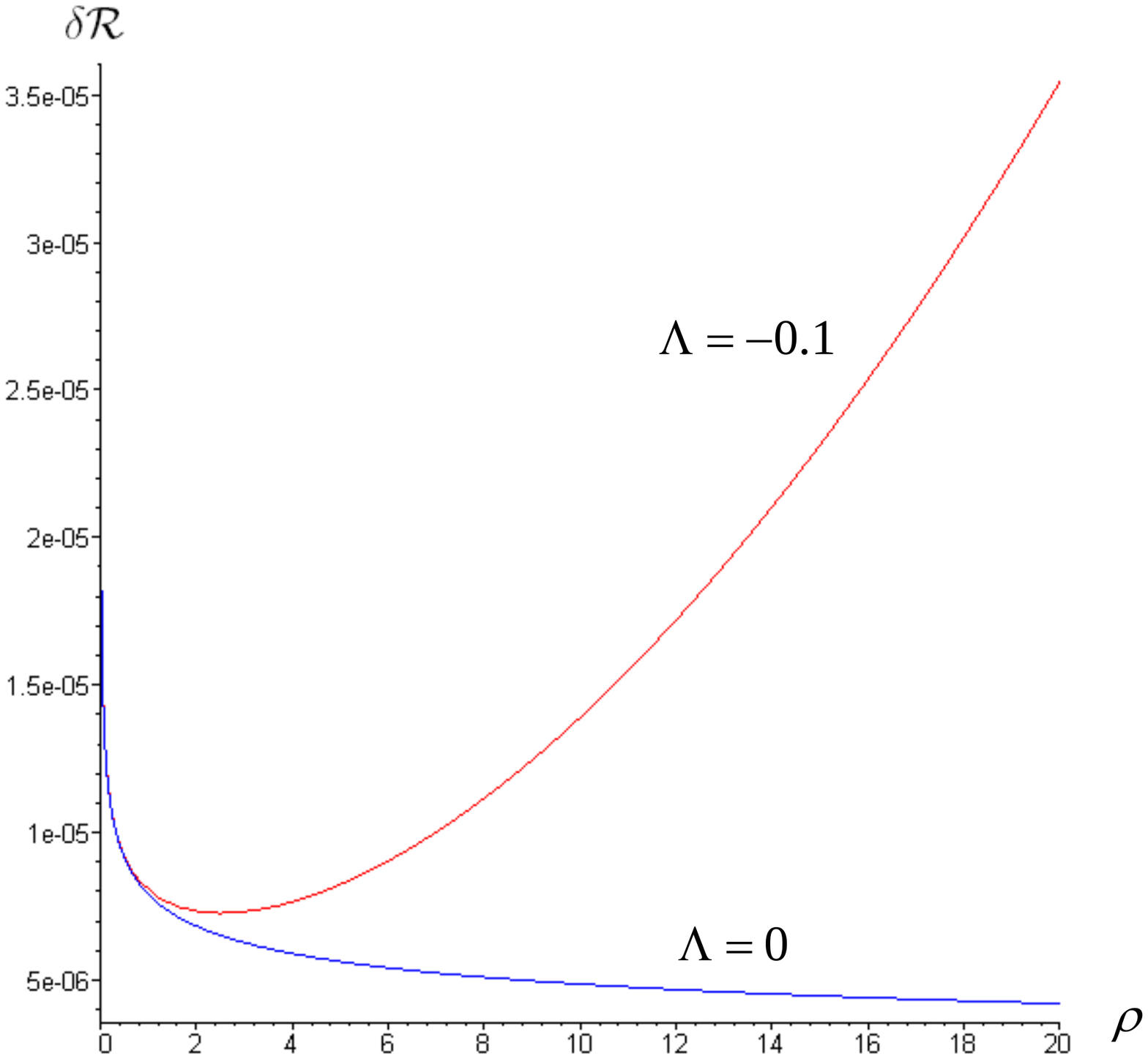}\\
\end{tabular}
\end{minipage}
\caption{ \textsl{Left panel}: Graphs of the proper radius $\mathcal{R}$ for $c =1$, $\sigma=1/5$, in the cases
$\Lambda=0$, where $\mathcal{R}=\sqrt{\rho^{2(1-2\sigma)}},$ and $\Lambda=-0.1$. \textsl{Right panel}: Graphs of $\delta\mathcal{R}$ for $\Lambda=0$ and $\Lambda=-0.1$, considering $\delta\rho =0.00001$. These figures illustrate $\Lambda$ destabilising $\mathcal{R}$ and $\delta\mathcal{R}$.}
\label{fig:pR}
\end{figure}

We study the proper radius $\mathcal{R}=\sqrt{g_{33}(\rho)}$, as measured in the LT spacetime, under small finite changes $\delta\rho$ in the radial coordinate $\rho$ given by
\begin{equation}
\delta\mathcal{R}=\sqrt{g_{33}(\rho+\delta\rho)}-\sqrt{g_{33}(\rho)}. \label{24}
\end{equation}
From (\ref{four}) and (\ref{5}), we obtain
\begin{eqnarray}
Q(\rho+\delta\rho)=Q(\rho)\left[1+\sqrt{3|\Lambda|}\frac{\delta\rho}{\tanh(2R)}\right], \label{twentyfive}\\
P(\rho+\delta\rho)=P(\rho)\left[1+\sqrt{3|\Lambda|}\frac{\delta\rho}{\sinh(2R)}\right], \label{26}
\end{eqnarray}
following
\begin{eqnarray}
\sqrt{g_{33}(\rho+\delta\rho)}=\sqrt{g_{33}(\rho)}\left\{1+\sqrt{\frac{|\Lambda|}{3}}
\left[\cosh(2R)
+\frac{2(1-2\sigma-2\sigma^2)}{\Sigma}\right]\frac{\delta\rho}{\sinh(2R)}\right\}. \label{27}
\end{eqnarray}
Substituting (\ref{27}) into (\ref{24}) we obtain
\begin{eqnarray}
\delta\mathcal{R}=
c\left(\frac{3|\Lambda|}{4}\right)^{2\sigma^2/\Sigma}
\frac{(\tanh R)^{4(1-\sigma)^2/3\Sigma}}{(\sinh R)^{4/3}}
\left(\frac{2}{3}\sinh^2R+\frac{1-2\sigma}{\Sigma}\right)\delta\rho, \label{28}
\end{eqnarray}
which shows that, with increasing $\rho$, the proper radius increases too (see also \cite{Zofka-Bicak}). Examples of this dynamical behaviour are depicted in Figure \ref{fig:pR}.

For the LC spacetime, $|\Lambda|=0$, (\ref{28}) reduces to (see also
\cite{Griffiths-Book} p. 176)
\begin{equation}
\delta\mathcal{R}_{LC}=\frac{c}{\Sigma}(1-2\sigma)\rho^{-4\sigma^2/\Sigma}\delta\rho, \label{29}
\end{equation}
from which can be seen that for $\sigma=0$, $\delta\mathcal{R}_{LC}=c\delta\rho$ and for $\sigma=1/2$, $\delta\mathcal{R}_{LC}=0$, as expected.

From (\ref{29}), we have that the variation of the proper radius diminishes gradually, while the coordinate $\rho$ increases,
and when $\rho\rightarrow\infty$ we have $\delta{\mathcal R}_{LC}\rightarrow 0$. The fact that for large distances $\rho$, the
 proper radius becomes nearly constant might explain, in part,
  the fact that the LC tangential velocity does not depend upon $\rho$.

If $|\Lambda|\ll1$, then up to first order, (\ref{28}) becomes
\begin{eqnarray}
\delta{\mathcal R}\approx\delta{\mathcal R}_{LC}+|\Lambda|\frac{c}{\Sigma^2}\;\sigma^2[3(1-2\sigma)+8\sigma^2]
\rho^{2(1-2\sigma+2\sigma^2)/\Sigma}
\delta\rho, \label{30}
\end{eqnarray}
where we see that the contribution due to $|\Lambda|$ increases the proper distance along $\rho$. Furthermore, from the exact expression ${\mathcal R}=\sqrt{g_{33}}$ we get that, as $\rho\to\infty$, ${\mathcal R}$ grows as $\exp(\sqrt{3|\Lambda|}\rho/6)$. In this sense, the proper radius of circular orbits in static vacuum cylindrically symmetric LC spacetimes is asymptotically unstable to the introduction of $\Lambda<0$.

We will get back to the analysis of circular  orbits in Section 5.1, where we calculate  the minimum radius of the stable orbits.

\section{Geodesics along $z$ ($\dot \phi=0$)}
\label{sec-along-z}
In the case ${\dot\phi}=0$, we have from (\ref{9}) and (\ref{13})
\begin{eqnarray}
\dot z=P_z\frac{P^{2\zeta/3}}{Q^{2/3}}\\
{\ddot z}=\frac{2}{3}P_z
\left[\zeta-\cosh(2R)\right]\frac{P^{2\zeta/3}}{Q^{5/3}}{\dot\rho}, \label{32}
\end{eqnarray}
where
\begin{equation}
\zeta(\sigma)=\frac{1+4\sigma-8\sigma^2}{\Sigma}. \label{34}
\end{equation}
We start by noting that for any $0\le \sigma\le 1/2$ and $P_z>0$ we have $\dot z>0$, for all $\rho\in~]0,\infty[$, showing that the geodesics along the $z$ direction are unbounded.

If $\Lambda=0$, then (\ref{32}) reduces to (see also \cite{Herrera2})
\begin{equation}
{\ddot z}=P_z\frac{4\sigma(1-2\sigma)}{\Sigma}
\frac{\dot\rho}{\rho^{(1-6\sigma+12\sigma^2)/\Sigma}}, \label{33}
\end{equation}
implying that, for $0<\sigma<1/2$ and $P_z>0$, the particle along the $z$ direction tends always to accelerate (decelerate) for $\dot\rho>0$ ($\dot\rho<0$), i.e. for increasing (decreasing) radial distances from the axis.
For $\rho\rightarrow 0$ we have, from (\ref{32}), the same behaviour (\ref{33}) since the spacetime LT tends to the LC spacetime. However, for $\sigma=0,1/2$ one gets ${\ddot z}=0$.

If $\Lambda< 0$, the dynamics along the $z$ direction is more complex than in the $\Lambda=0$ case described above.
For $\Lambda<0$, from (\ref{34}), we have
\begin{equation}
\frac{d\zeta}{d\sigma}=\frac{6(1-4\sigma)}{\Sigma^2}, \label{35}
\end{equation}
which shows that $\zeta(\sigma)$ has a maximum at $\sigma=1/4$
attaining $\zeta=2$, while for $\sigma=0, 1/2$ becomes
$\zeta=1$.

\begin{figure}[H]
\begin{minipage}{8cm}
\begin{tabular}{c}
\includegraphics[width=8cm]{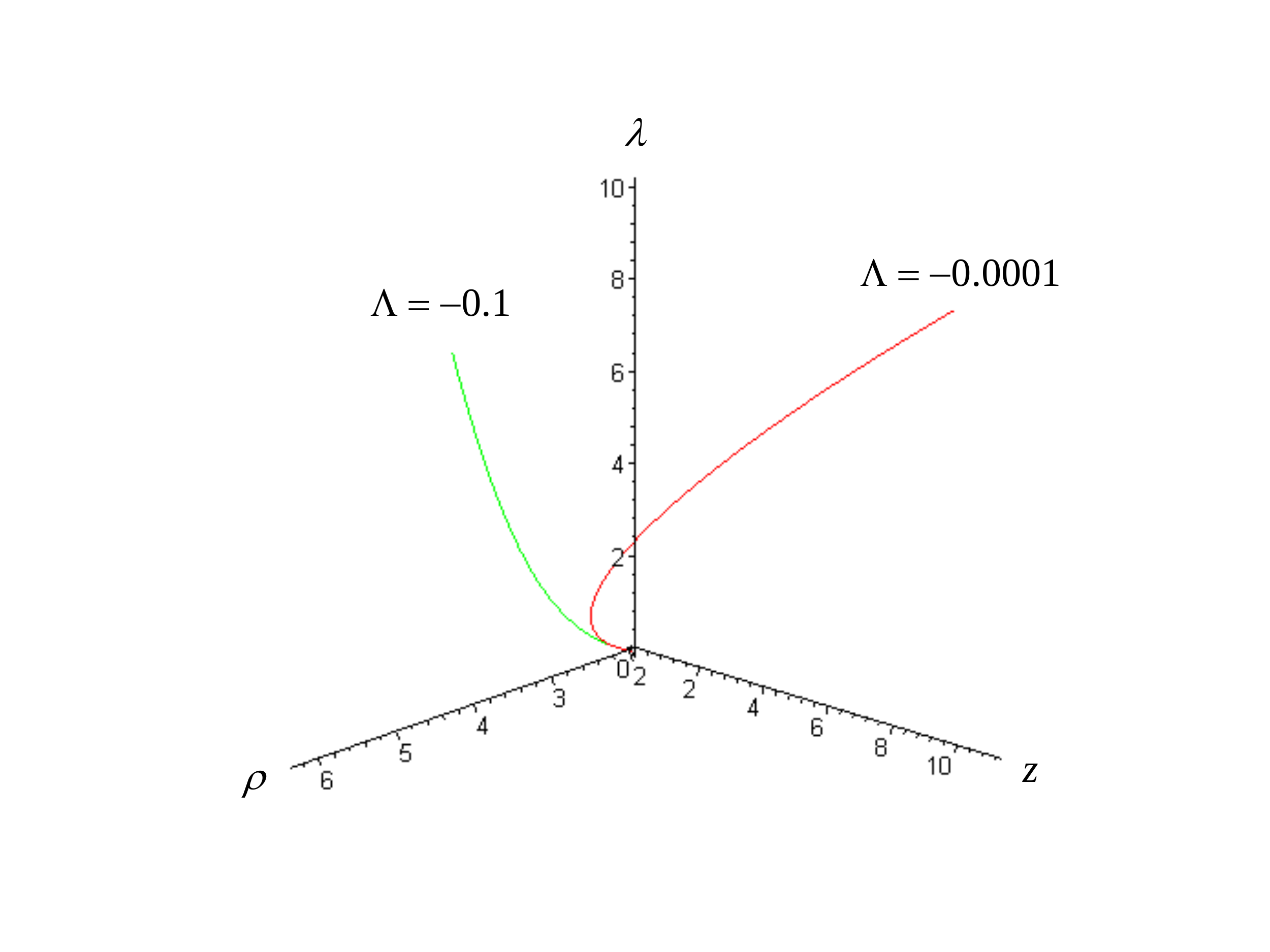} \\
\end{tabular}
\end{minipage}
\begin{minipage}{8cm}
\begin{tabular}{c}
\includegraphics[width=8cm]{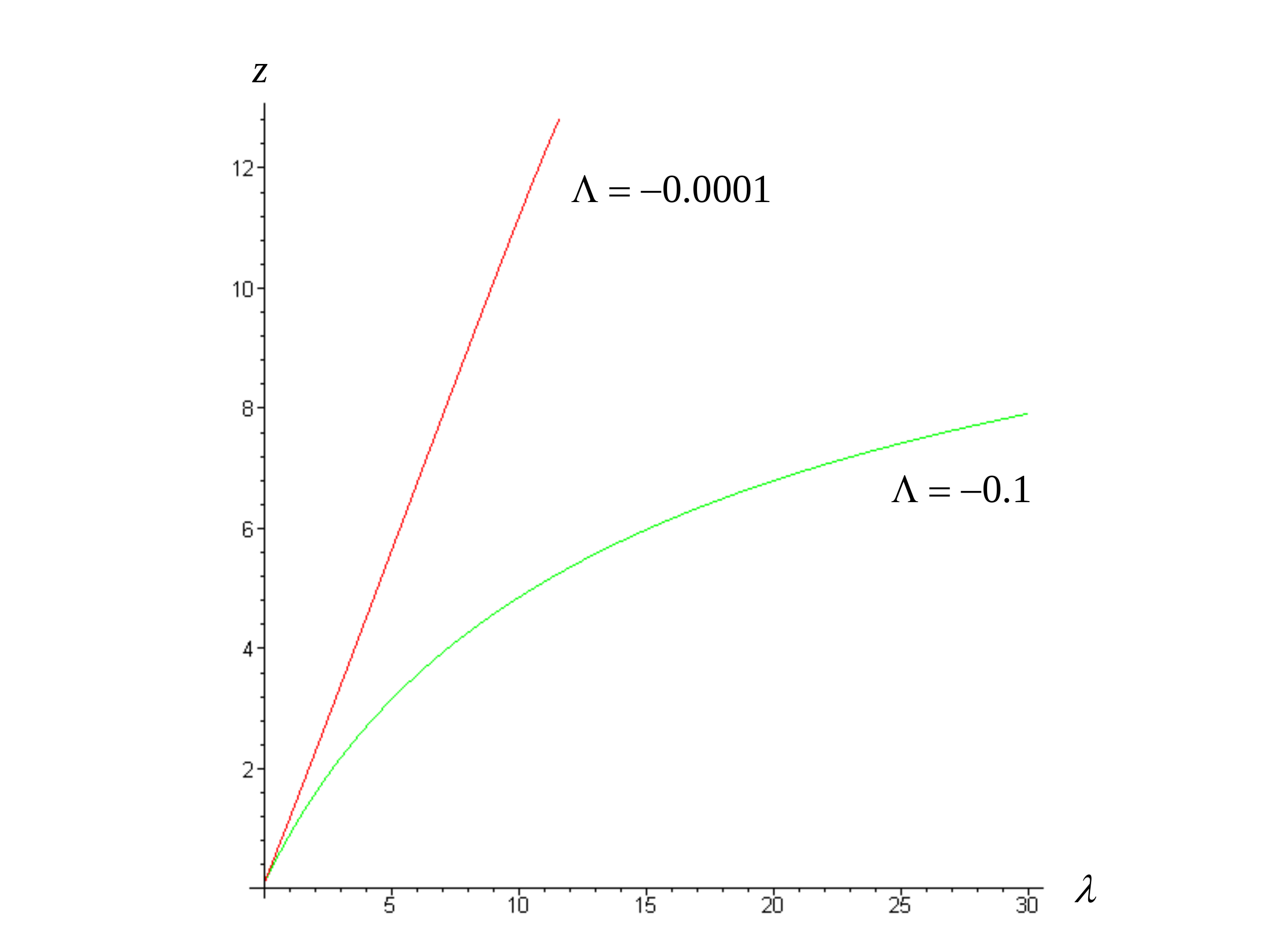}\\
\end{tabular}
\end{minipage}
\caption{ Graphs of the numerical integration of the geodesics' equations along $z(\lambda)$, for $E=2$, $P_z=c=1$, $L_z =\epsilon =0$, $\sigma=1/5$, $\lambda\in[0,30]$ on the right and $\lambda\in[0,10]$ on the left for $\Lambda=-0.1$ and for $\Lambda=-0.0001$. }
\label{fig:geodz}
\end{figure}

For $0<\sigma<1/2$ and $P_z>0$, when a particle is moving close to the
$\rho=0$ axis, $\cosh(2R)<\zeta$, and is radially distancing from
the axis, $\dot\rho>0$, its acceleration along the $z$ direction is
positive and it attains its maximum speed for $\cosh(2R)=\zeta$.
While the particle continues distancing radially, for distances
$\cosh{(2R)}>\zeta$, the particle speed along $z$ diminishes, ${\ddot
z}<0$, and tends to zero as $\rho\rightarrow\infty$. Numerical examples of this behaviour are depicted in Figure \ref{fig:geodz}. 

This effect is
reversed for particles approaching radially the axis, $\dot\rho<0$,
and for large radial distances, $\cosh(2R)>\zeta$. In this case, the particle
acceleration along $z$ is ${\ddot z}>0$, the particle attains
its maximum speed along $z$ for $\cosh(2R)=\zeta$, which, then,
 gradually diminishes while tending to zero when $\rho\rightarrow 0$.

For $\sigma=0$ or $\sigma=1/2$, we have from (\ref{32}) and (\ref{13})
\begin{eqnarray}
{\ddot z}=\frac{2}{3}P_z\left[1-\cosh(2R)\right]\frac{P^{2/3}}{Q^{5/3}}{\dot\rho}, \label{36}\\
{\dot z}=\frac{P_z}{(\cosh R)^{4/3}}, \label{37}
\end{eqnarray}
 which shows that both the acceleration and the velocity decrease and tend to zero as $\rho\rightarrow\infty$. While, if $\Lambda=0$, then (\ref{36}) and (\ref{37}) reduce to ${\ddot z}=0$ and
${\dot z}=P_z$.

We stress that the geodesics' motions described above stem purely from General Relativity and have no Newtonian analog.

It is interesting to observe that geodesics along the axis in the van Stockum spacetime \cite{Opher}, describing a rigidly rotating dust, have a similar behaviour as in the LC spacetime \cite{Herrera2}. The inclusion of the cosmological constant in the van Stockum spacetime, producing the Lanczos spacetime, shows no important differences in the geodesics' dynamics along the axis, as compared to the van Stockum counterpart \cite{Pereira}. However, as we have seen in this section, the LT spacetime modifies dramatically the behaviour of geodesics along the axis as compared to the results that emerge from the LC spacetime.

\section{Geodesics along $\rho$}

In order to analyse the geodesics' dynamics along $\rho$, we introduce a
potential $V(\rho)$ of the form
\begin{equation}
V(\rho)=\epsilon
Q^{2/3}P^{-2(1-8\sigma+4\sigma^2)/3\Sigma}+P_z^2P^{8\sigma(1-\sigma)/\Sigma}+\left(\frac{L_z}{c}\right)^2
P^{-2(1-4\sigma)/\Sigma} \label{39}
\end{equation}
which is always positive for null ($\epsilon=0$) and timelike
($\epsilon=1$) geodesics. As $\rho\rightarrow 0$, we get the following useful asymptotic estimates:
$V\rightarrow\infty$, for $0<\sigma<1/4$;  $V\rightarrow 0$, for
$1/4<\sigma\le1/2$; $V\rightarrow (L_z/c)^2$, for $\sigma=1/4$; $V\to \infty$, for $\sigma=0$ and $L_z\ne 0$;  $V\to \epsilon+P_z^2$, for $\sigma=L_z=0$. Estimates for $V$ as $\rho\to\infty$ depend on $\epsilon$ and will be given ahead, in each case.

The above potential will play an important role in the next sections and its definition is motivated by the fact that, from (\ref{12}), we can write $\dot \rho^2$ as
\begin{equation}
{\dot\rho}^2=[E^2-V(\rho)]Q^{-2/3}P^{2(1-8\sigma+4\sigma^2)/3\Sigma}.
\label{38}
\end{equation}
Since $P$ and $Q$ are finite and non-zero for $\rho\in~]0,\infty[$ then, $\dot \rho=0$ if and only if $V(\rho)=E^2$.

The equation $V(\rho)=E^2$ allows to find the minimum or maximum distances, $\rho=\rho_{min}$ or $\rho=\rho_{max}$, when they exist, reached by a particle from the axis. In the case $|\Lambda|=0$, that equation reduces to
\begin{equation}
\epsilon \rho_{LCm}^{4\sigma/\Sigma}+ P_{z}^{2} \rho_{LCm}^{8\sigma(1-\sigma)/\Sigma}+\left(\frac{L_z}{c}\right)^2 \rho_{LCm}^{-2(1-4\sigma)/\Sigma}=E^2\label{38A}
\end{equation}
and, up to first order in $|\Lambda|$, gives
\begin{eqnarray}
\rho_{m}\approx\rho_{LCm} +\;\frac{|\Lambda|}{4}\; \left[\left(\frac{L_z}{c}\right)^2 (1-4\sigma) \rho_{LCm}^{4\sigma(1+2\sigma)/\Sigma}\right. \nonumber\\
\left.+ \epsilon (1-2\sigma)^2 \rho_{LCm}^{2(1+4\sigma^2 )/ \Sigma}- 4P_{z}^2 \sigma(1-\sigma)\rho_{LCm}^{2(1+2\sigma)/\Sigma} \right] \nonumber\\
\times
\left[\left(\frac{L_z}{c}\right)^2 (1-4\sigma) \rho_{LCm}^{(-3+10\sigma-4\sigma^2)/\Sigma} -2\epsilon \sigma \rho_{LCm}^{(-1+6\sigma-4\sigma^2)/\Sigma} \right.\nonumber\\
\left.-4P_{z}^{2}\sigma(1-\sigma)\rho_{LCm}^{(-1+10\sigma-12\sigma^2)/\Sigma} \right]^{-1},\label{38B}
\end{eqnarray} 
where $\rho_{LCm}$ and $\rho_{m}$ denote the extreme (minimum or maximum) values of $\rho$ in the LC and LT spacetimes, respectively. From this relation, we
will be able to conclude, in some cases, that $|\Lambda|$ increases the extreme distances of the geodesics to the axis. Although (\ref{38B}) is approximate, we will obtain an exact relation between $\rho_m$ and $\rho_{LCm}$ for the case $\epsilon=P_z=0$, in Section 5.1.1.

For a particle in the LC spacetime, (\ref{38}) becomes
\begin{equation}
{\dot\rho}^2_{LC}=\left[E^2- \epsilon \rho^{4\sigma/\Sigma}-P_{z}^{2} \rho^{8\sigma(1-\sigma)/\Sigma}-\left(\frac{L_z}{c}\right)^2 \rho^{-2(1-4\sigma)/\Sigma} \right]\rho^{-4\sigma/\Sigma},
\label{38C}
\end{equation}
and considering $|\Lambda|$ small, we obtain from (\ref{38}) and (\ref{38C}), up to first order in $|\Lambda|$,
\begin{equation}
{\dot\rho}^2\approx{\dot\rho}^2_{LC}+\frac{|\Lambda|}{2\Sigma}\left[4\sigma^2\left(\frac{L_z}{c}\right)^2\rho^{8\sigma^2/\Sigma}
-(1-2\sigma)^2E^2\rho^{2(1-2\sigma)^2/\Sigma}+P_{z}^{2} \rho^{2/\Sigma}\right].
\end{equation}
Since the extremes of the potential may correspond to limit orbits, we compute its derivatives from (\ref{39}) as
\begin{eqnarray}
V^{\star}(\rho)=\frac{2\sqrt{3|\Lambda|}}{\Sigma \sinh(2R)}
\left\{\frac{\epsilon}{3} \left[\Sigma
\cosh(2R)-1+8\sigma-4\sigma^2\right]Q^{2/3}P^{-2(1-8\sigma+4\sigma^2)/3\Sigma}
\right. \nonumber\\
\left.+4\sigma(1-\sigma)P^2_zP^{8\sigma(1-\sigma)/\Sigma}-
(1-4\sigma)\left(\frac{L_z}{c}\right)^2P^{-2(1-4\sigma)/\Sigma}\right\}
\label{40}
\end{eqnarray}
and
\begin{eqnarray}
V^{\star\star}(\rho)=-\frac{\sqrt{3|\Lambda|}V^{\star}}{\tanh(2R)} 
+\left[\frac{2\sqrt{3|\Lambda|}}{\Sigma \sinh(2R)}\right]^2
\left\{\frac{\epsilon}{6}\Sigma^2Q^{2/3}P^{-2(1-8\sigma+4\sigma^2)/3\Sigma}[\sinh(2R)]^2\right.\nonumber\\
\left.+\frac{\epsilon}{9}Q^{2/3}P^{-2(1-8\sigma+4\sigma^2)/3\Sigma}\left[\Sigma \cosh(2R)-1+8\sigma-4\sigma^2\right]^2\right. \nonumber\\
\left.+16\sigma^2(1-\sigma)^2P_z^2P^{8\sigma(1-\sigma)/\Sigma}+(1-4\sigma)^2\left(\frac{L_z}{c}\right)^2
P^{-2(1-4\sigma)/\Sigma}\right\}. \label{41}
\end{eqnarray}
We note that $V^\star$ does not depend on $L_z$ for $\sigma=1/4$, and it does not depend on $P_z$ for $\sigma=0$. This will help to clarify ahead the geodesics's dynamics in those cases.

Whenever the equation $V^{\star}(\rho)=0$ has a solution, say $\rho=\rho_e$,
satisfying
\begin{eqnarray}
(1-4\sigma)\left(\frac{L_z}{c}\right)^2P^{-2(1-4\sigma)/\Sigma} 
\stackrel{\rho_e}{=}
\frac{\epsilon}{3}\;Q^{2/3}P^{-2(1-8\sigma+4\sigma^2)/3\Sigma}
\left[\Sigma \cosh(2R)-1+8\sigma-4\sigma^2\right] \nonumber\\
+4\sigma(1-\sigma)P_z^2P^{8\sigma(1-\sigma)/\Sigma}, \label{42}
\end{eqnarray}
where $\stackrel{\rho_e}{=}$ denotes evaluation at $\rho=\rho_e$,
then by substituting into (\ref{26}) we obtain,
\begin{eqnarray}
V^{\star\star}(\rho_e)=\left[\frac{2\sqrt{3|\Lambda|}}{\Sigma
\sinh(2R)}\right]^2\left\langle\frac{\epsilon}{3}Q^{2/3}
P^{-2(1-8\sigma+4\sigma^2)/3\Sigma}\left\{\frac{\Sigma^2}{2}[\sinh(2R)]^2\right.\right.\nonumber\\
\left.\left.+\frac{1}{3}\left[\Sigma \cosh(2R)-1+8\sigma
-4\sigma^2\right]^2+(1-4\sigma)\left[\Sigma \cosh(2R)-1+8\sigma-4\sigma^2\right]\right\}\right.\nonumber\\
\left.+4\sigma(1-4\sigma^2)(1-\sigma)P_z^2P^{8\sigma(1-\sigma)/\Sigma}\right\rangle .
\label{43}
\end{eqnarray}
Finally, from (\ref{8}) with (\ref{11}), (\ref{13}) and
(\ref{14}), we get for the acceleration
\begin{eqnarray}
{\ddot\rho}=-\frac{P^{2(1-8\sigma+4\sigma^2)/3\Sigma}}{3\Sigma
Q^{5/3}}\left[\Sigma\cosh(2R)-1+8\sigma-4\sigma^2\right]
\nonumber\\
\times
\left\{E^2-P_z^2\left[\frac{\Sigma\cosh(2R)-1-4\sigma+8\sigma^2}{\Sigma
\cosh(2R)-1+8\sigma-4\sigma^2}\right]
P^{8\sigma(1-\sigma)/\Sigma}\right. \nonumber\\
-\left.\left(\frac{L_z}{c}\right)^2\left[\frac{\Sigma\cosh(2R)+2(1-2\sigma-2\sigma^2)}
{\Sigma\cosh(2R)-1+8\sigma-4\sigma^2}\right]P^{-2(1-4\sigma)/\Sigma}\right\},
\label{44}
\end{eqnarray}
which will also be used ahead.

We now split the analysis into planar and non-planar as well as null and non-null geodesics, and use the general formulae (\ref{39})-(\ref{44}) to study the geodesics' dynamics, in each case.

\subsection{Planar Geodesics ($\dot z=0$)}
\label{sec-planar}

Planar geodesics, with $P_z=0$, for $\Lambda<0$, were analysed in \cite{Banerjee} where they found that, depending on some constants related to $\sigma$, some families of null and timelike geodesics may be trapped. Here, we study those aspects in more detail using the above defined potential which clarifies the physical meaning of some constants of  \cite{Banerjee}. We also look more deeply at the impact of $\Lambda$ on the existence of geodesics confinement and, in particular, on their minimum and/or maximum possible radii.

\subsubsection{Case $\epsilon=0$}

In this case, (\ref{39}) becomes
\begin{equation}
V(\rho)=\left(\frac{L_z}{c}\right)^2P^{-2(1-4\sigma)/\Sigma}, \label{45}
\end{equation}
and, if $\sigma=1/4$, it becomes constant as
\begin{equation}
V_{1/4}=\left(\frac{L_z}{c}\right)^2. \label{45aa}
\end{equation}
The asymptotic behaviour, $\rho\rightarrow\infty$, of (\ref{45}) is $V\rightarrow V_{\infty}$ where
\begin{equation}
V_{\infty}=\left(\frac{2}{\sqrt{3|\Lambda|} }\right)^{-2(1-4\sigma)/\Sigma}\left(\frac{L_z}{c}\right)^2. \label{45a}
\end{equation}
Furthermore, we have $V^{\ast}(\rho)<0$ if $\sigma<1/4$, $V^{\ast}(\rho)=0$ if $\sigma=1/4$ and $V^{\ast}(\rho)>0$ if $\sigma>1/4$, suggesting that we can separate our analysis into three different cases, as follows:
\begin{enumerate}
\item{$\sigma<1/4$}
\begin{enumerate}
\item If {$E^2>V_{\infty}$}  and $L_z\ne 0$, from (\ref{38}) and (\ref{44}), a null particle approaching $z$ has decreasing negative acceleration, $\ddot\rho<0$, and increasing speed $\dot\rho$ attaining its maximum speed at $\ddot\rho=0$. From this point, its speed diminishes since $\ddot\rho>0$, and the particle reaches its minimum distance from the axis for
    \begin{equation}
    P_{min}=\left(\frac{L_z}{cE}\right)^{\Sigma/(1-4\sigma)}, \label{46a}
    \end{equation}
    from which we can extract $\rho_{min}$. At $\rho_{min}$, the null particle is reflected to infinity, $\rho\rightarrow\infty$, where $\dot\rho\rightarrow 0$. For $|\Lambda|=0$, we have from (\ref{46a})
    \begin{equation}
    \rho_{LCmin}=\left(\frac{L_z}{cE}\right)^{\Sigma/(1-4\sigma)}, \label{46b}
    \end{equation}
    which is the minimum distance from the $z$ axis attained by an incoming null particle in the LC spacetime.
    From (\ref{46a}) and (\ref{46b}), we have
    \begin{equation}
    \frac{\sqrt{3|\Lambda|}}{2}\,\rho_{LCmin}=\tanh\left(\frac{\sqrt{3|\Lambda|}}{2}\,\rho_{min}\right), \label{46bb}
    \end{equation}
    implying $\rho_{min}\geq\rho_{LCmin}$, which shows that $|\Lambda|$ increases the minimum distance of the null particle to the $z$ axis. A linear version of this result, for small $|\Lambda|$, can immediately be derived from (\ref{38B}) and non-linear numerical examples are plotted in Figure \ref{fig:geod1}.

    For $V_{\infty}=0$ or $L_z=0$, incoming null particles hit the axis with infinite speed, $\dot\rho\rightarrow\infty$, while outgoing null particles escape to infinity, $\rho\rightarrow\infty$, attaining $\dot\rho\rightarrow 0$.

\item If $E^2\leq V_{\infty}$, then $\dot\rho^2<0$ which is physically not acceptable.
\end{enumerate}

\begin{figure}[H]
\begin{minipage}{8cm}
\begin{tabular}{c}
\includegraphics[width=8cm]{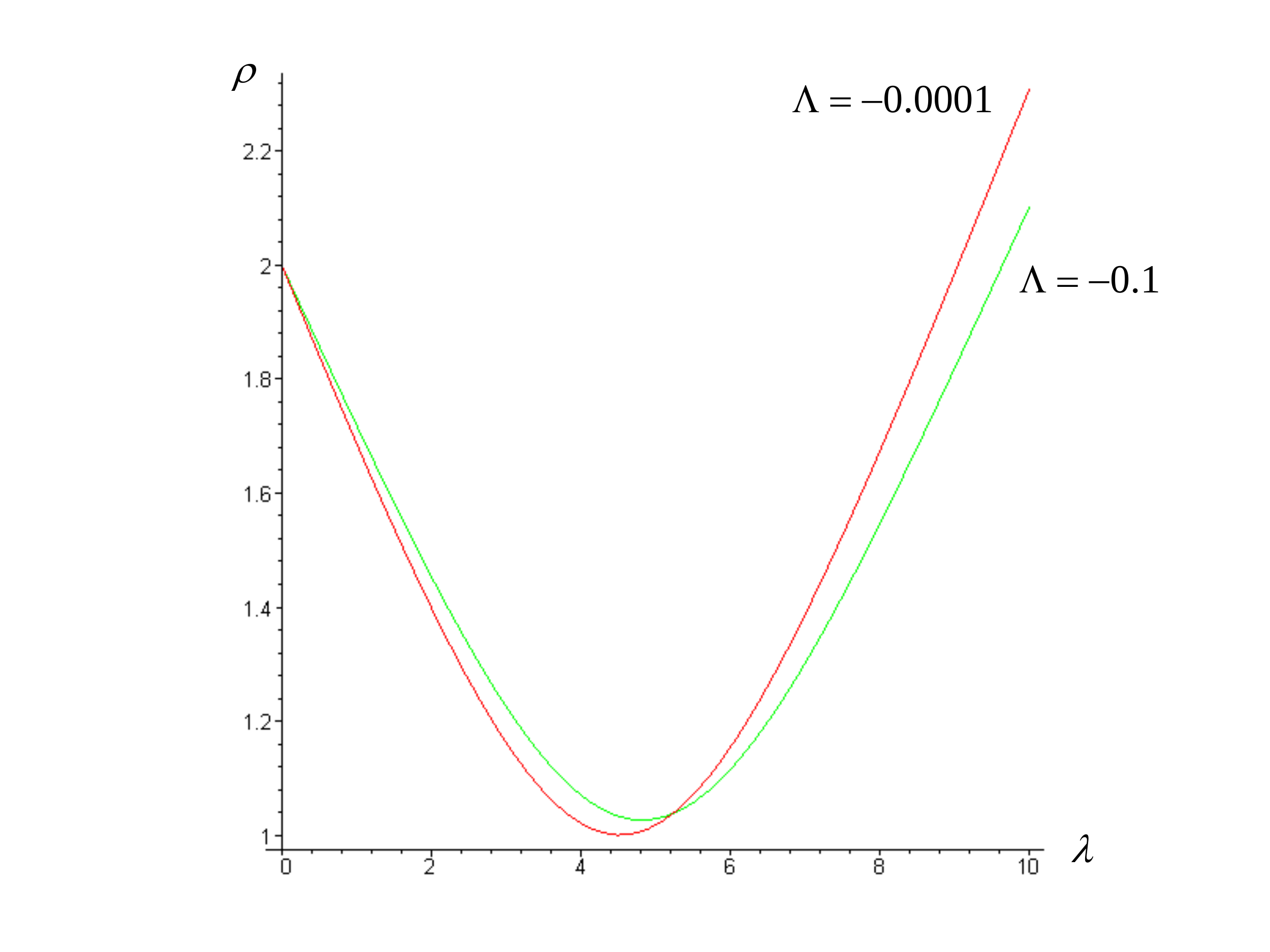} \\
\end{tabular}
\end{minipage}
\begin{minipage}{8cm}
\begin{tabular}{c}
\includegraphics[width=8cm]{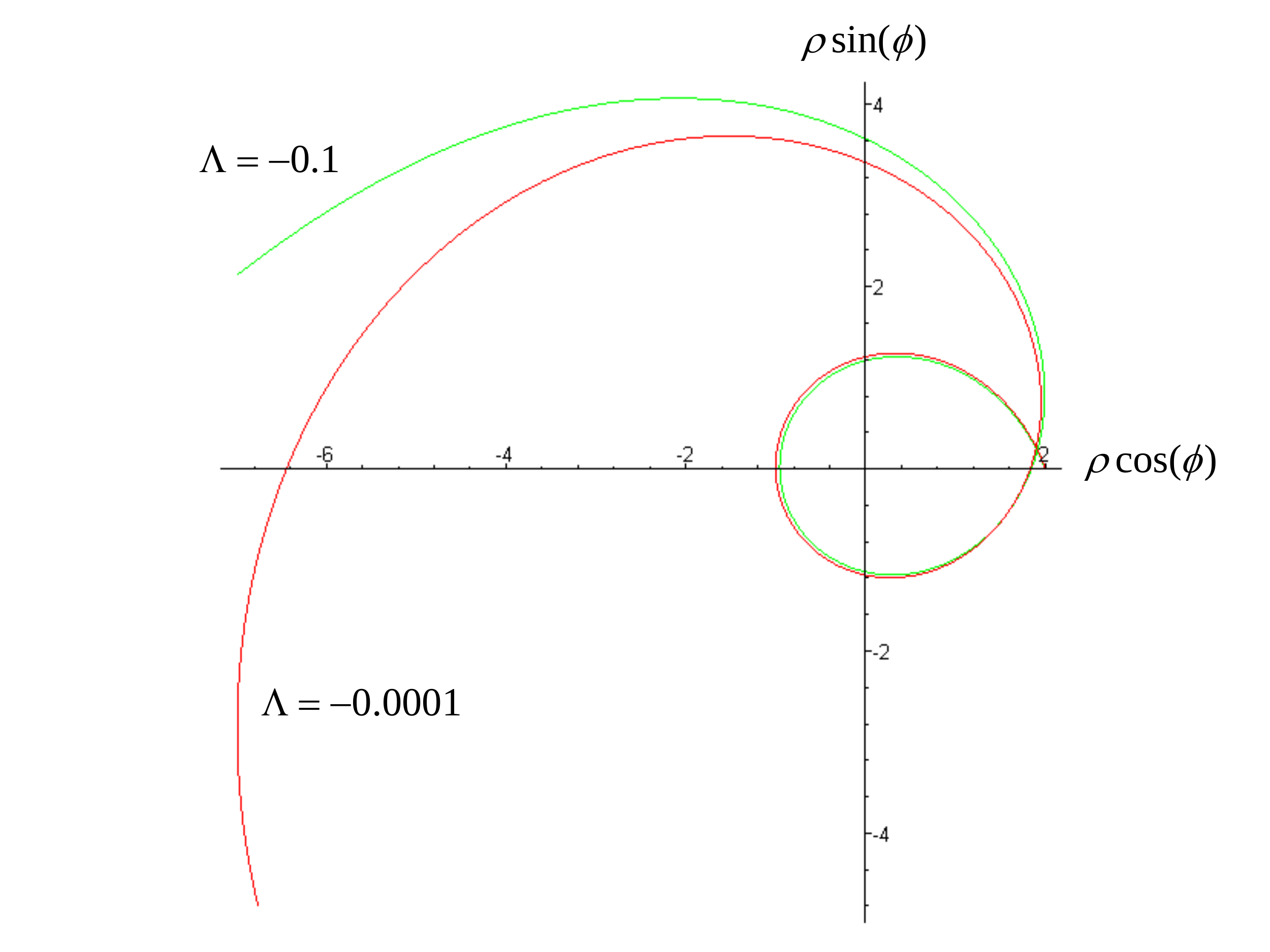}\\
\end{tabular}
\end{minipage}
\caption{ Graphs of the numerical integration of the geodesics' equations along $\rho(\lambda)$, for $E=L_z=c =1$, $P_z =\epsilon =0$, $\sigma=1/5$, $\lambda\in[0,10],$ in the cases
$\Lambda=-0.0001$ and $\Lambda=-0.1$, satisfying $E^2> V_{\infty}$. This illustrates the fact that, for $\epsilon=P_z=0$, increasing values of
$|\Lambda|$ increase the minimum distance, $\rho_{min}$, of null geodesics to the axis.}
\label{fig:geod1}
\end{figure}

\item{$\sigma=1/4$}
\begin{enumerate}
\item If $E^2>V_{1/4}$, incoming null particles hit the axis $z$ with infinite speed, $\dot\rho\rightarrow\infty$, as $\rho\rightarrow 0$, see examples in Figure \ref{fig:geod2}, while outgoing particles escape to infinity, $\rho\rightarrow\infty$, attaining $\dot\rho\rightarrow 0$. In particular, this also holds for $V_{1/4}=0$ or $L_z=0$.
\begin{figure}[H]
\begin{minipage}{8cm}
\begin{tabular}{c}
\includegraphics[width=8cm]{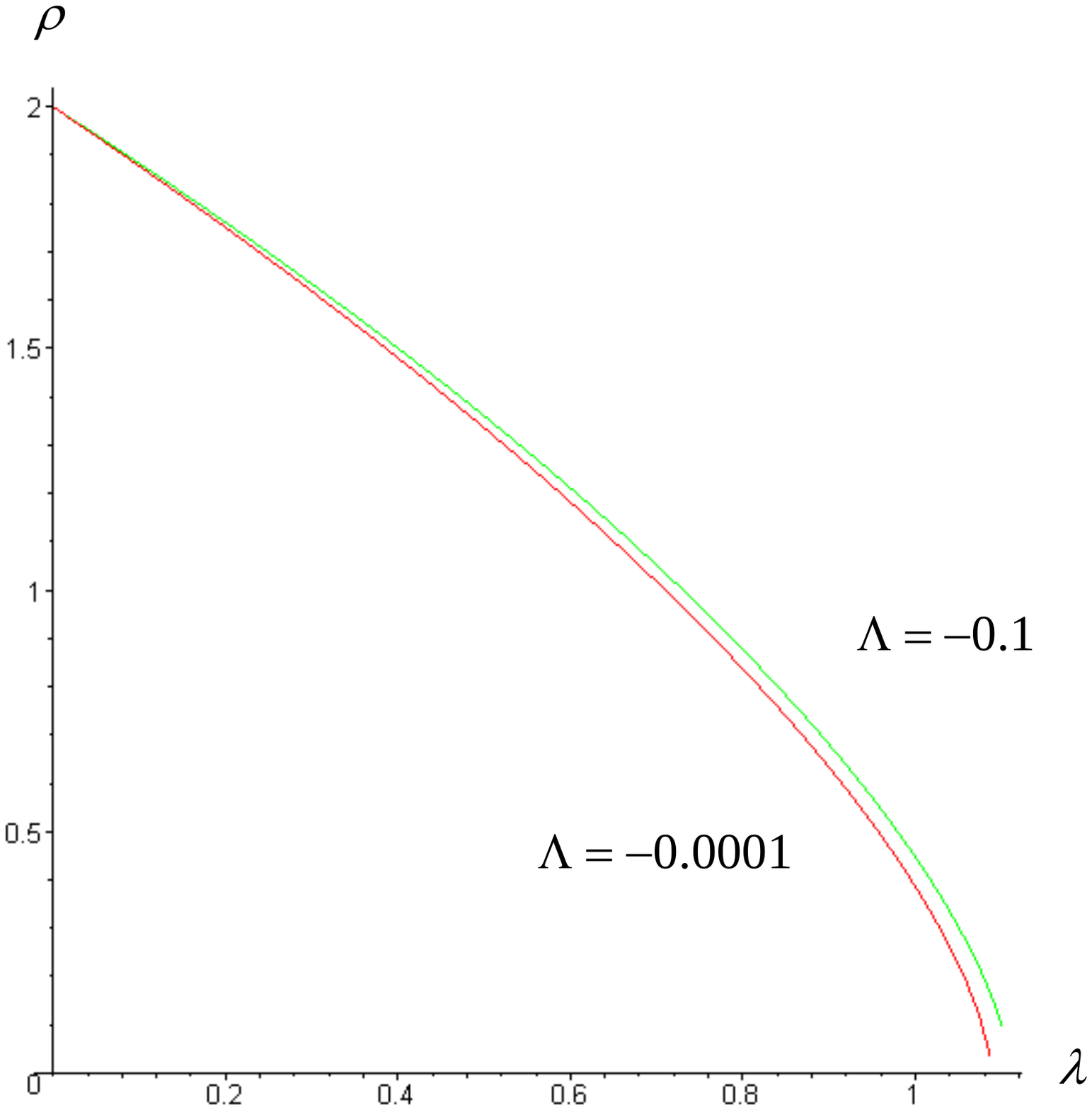} \\
\end{tabular}
\end{minipage}
\begin{minipage}{8cm}
\begin{tabular}{c}
\includegraphics[width=8cm]{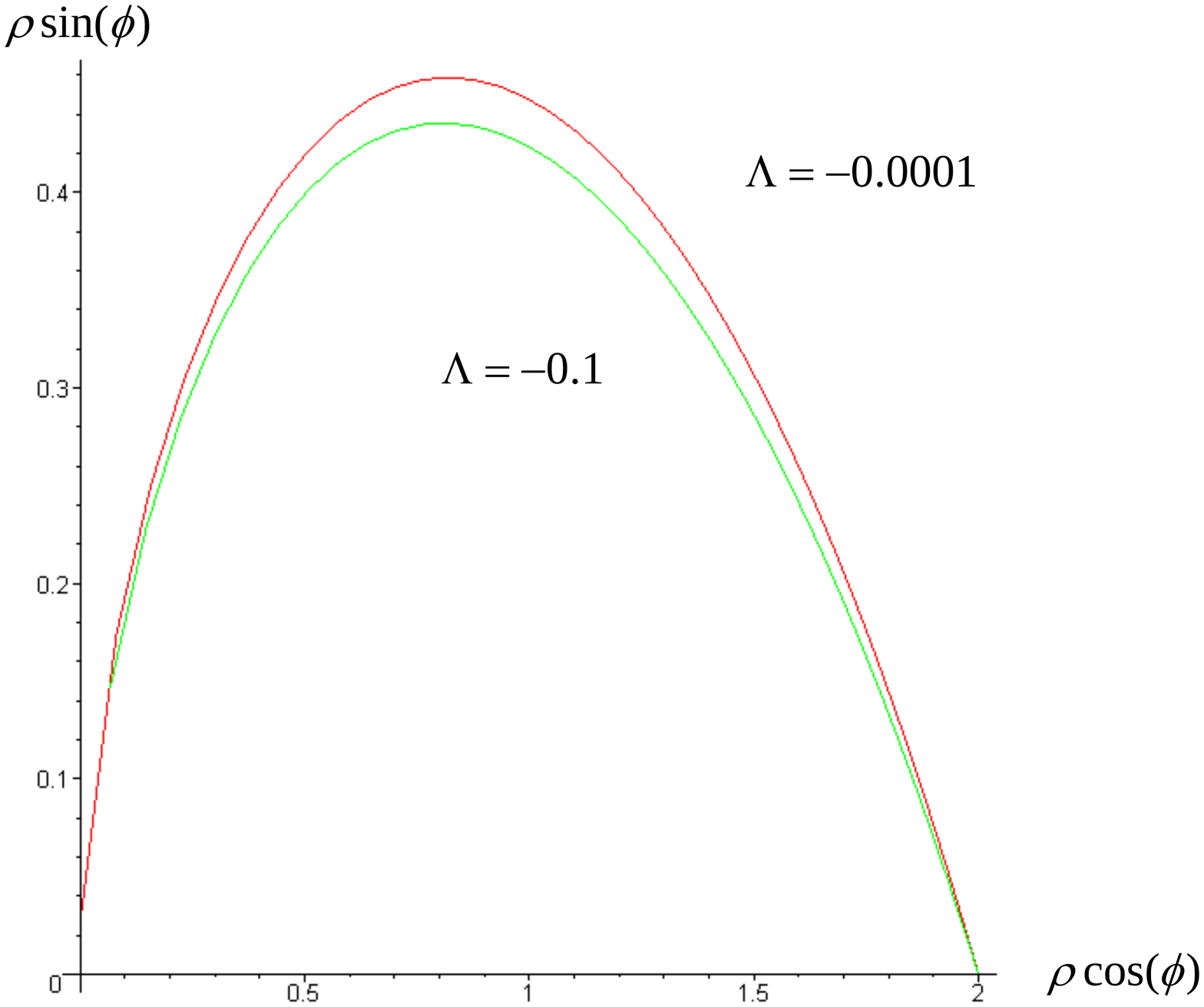}\\
\end{tabular}
\end{minipage}
\caption{ Graphs of the numerical integration of the geodesics' equations along $\rho(\lambda)$, for $E=2$, $P_z=\epsilon=0$, $L_z =c=1$, $\sigma=1/4$, in the cases
$\Lambda=-0.0001$ and $\Lambda=-0.1$, satisfying $E^2> V_{1/4}$.  }
\label{fig:geod2}
\end{figure}

\item If $E^2=V_{1/4}$ and $L_z\ne 0$, the radial speed of the null particle is zero, $\dot\rho=0$, and its motion is circular (see Section 3).
    \item If $E^2<V_{1/4}$, then $\dot\rho^2<0$ which is physically not acceptable.
\end{enumerate}

\item $\sigma>1/4$

\begin{enumerate}
\item If $E^2>V_{\infty}$, incoming null particles hit the axis $z$ with infinite speed, $\dot\rho\rightarrow\infty$, while outgoing particles escape to infinity, $\rho\rightarrow\infty$, attaining $\dot\rho\rightarrow 0$. See examples in Figure \ref{fig:geod3}.
 A similar behaviour holds for $V_{\infty}=0$ or $L_z=0$.

    \begin{figure}[H]
\begin{minipage}{8cm}
\begin{tabular}{c}
\includegraphics[width=8cm]{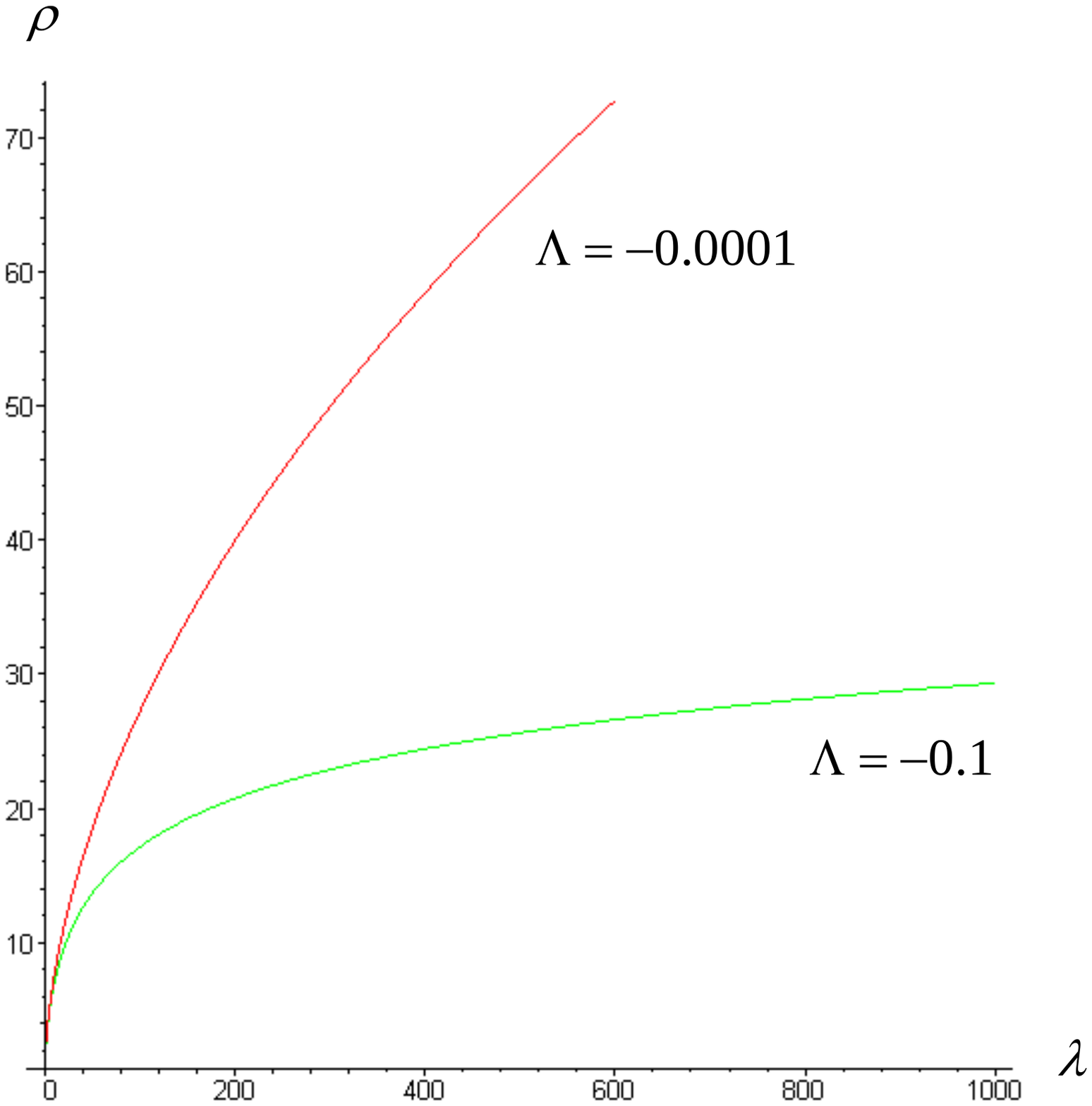} \\
\end{tabular}
\end{minipage}
\begin{minipage}{8cm}
\begin{tabular}{c}
\includegraphics[width=8cm]{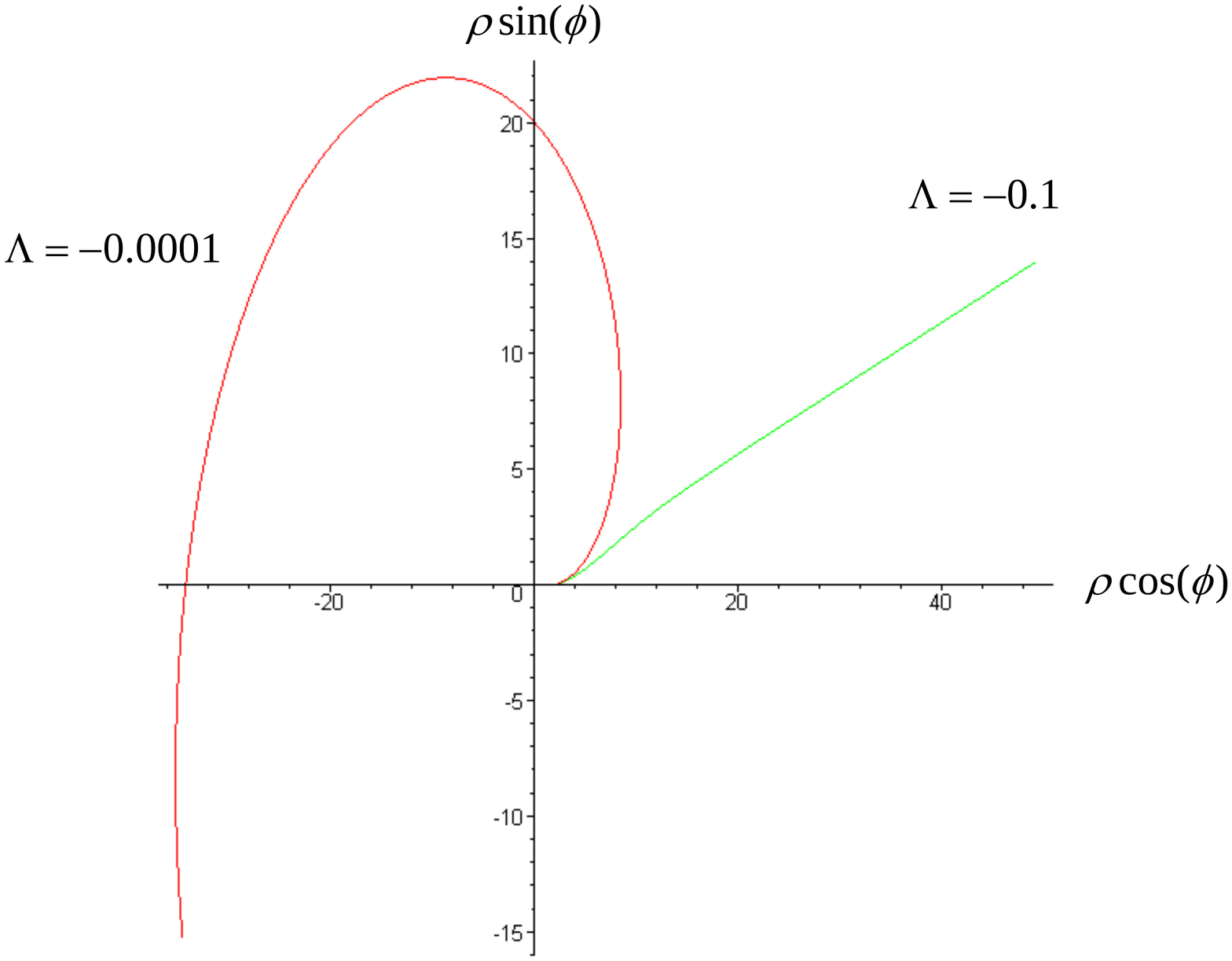}\\
\end{tabular}
\end{minipage}
\caption{ Graphs of the numerical integration of the geodesics' equations along $\rho(\lambda)$, for $E=2$, $P_z=\epsilon=0$, $L_z =0.1$, $c=1$, $\sigma=0.4$, in the cases
$\Lambda=-0.0001$ and $\Lambda=-0.1$, satisfying $E^2> V_{\infty}$.  }
\label{fig:geod3}
\end{figure}

\item If $E^2\leq V_{\infty}$  and $L_z\ne 0$, incoming null particles have increasing negative acceleration, $\ddot\rho<0$, and hit the axis with infinite speed $\dot\rho\rightarrow\infty$. However, from (\ref{38}) and (\ref{44}), outgoing null particles move with decreasing negative acceleration, $\ddot\rho<0$, and decreasing speed $\dot\rho$ attaining a maximum distance from the axis for
        \begin{equation}
        P_{max}=\left(\frac{cE}{L_z}\right)^{\Sigma/(4\sigma-1)}, \label{51a}
        \end{equation}
        from which we can extract $\rho_{max}$. At $\rho_{max}$, the null particle is reflected back to the axis attaining $\dot\rho\rightarrow\infty$. For $|\Lambda|=0$, we have from (\ref{51a})
        \begin{equation}
        \rho_{LCmax}=\left(\frac{cE}{L_z}\right)^{\Sigma/(4\sigma-1)}, \label{51b}
        \end{equation}
        which is the maximum distance from the $z$ axis attained by the outgoing null particle in the LC spacetime. From (\ref{51a}) and (\ref{51b}), we have
        \begin{equation}
        \frac{\sqrt{3|\Lambda|}}{2}\,\rho_{max}=\tanh\left(\frac{\sqrt{3|\Lambda| }}{2}\,\rho_{LCmax}\right), \label{51c}
        \end{equation}
        showing that $|\Lambda|$ increases the maximum distance to the axis $z$ reached by the null particle (see also Figure \ref{fig:geod4}).

         For $V_{\infty}=0$ or $L_z=0$, incoming null particles hit the axis $z$ while outgoing null particles escape to infinity.
\begin{figure}[H]
\begin{minipage}{8cm}
\begin{tabular}{c}
\includegraphics[width=8cm]{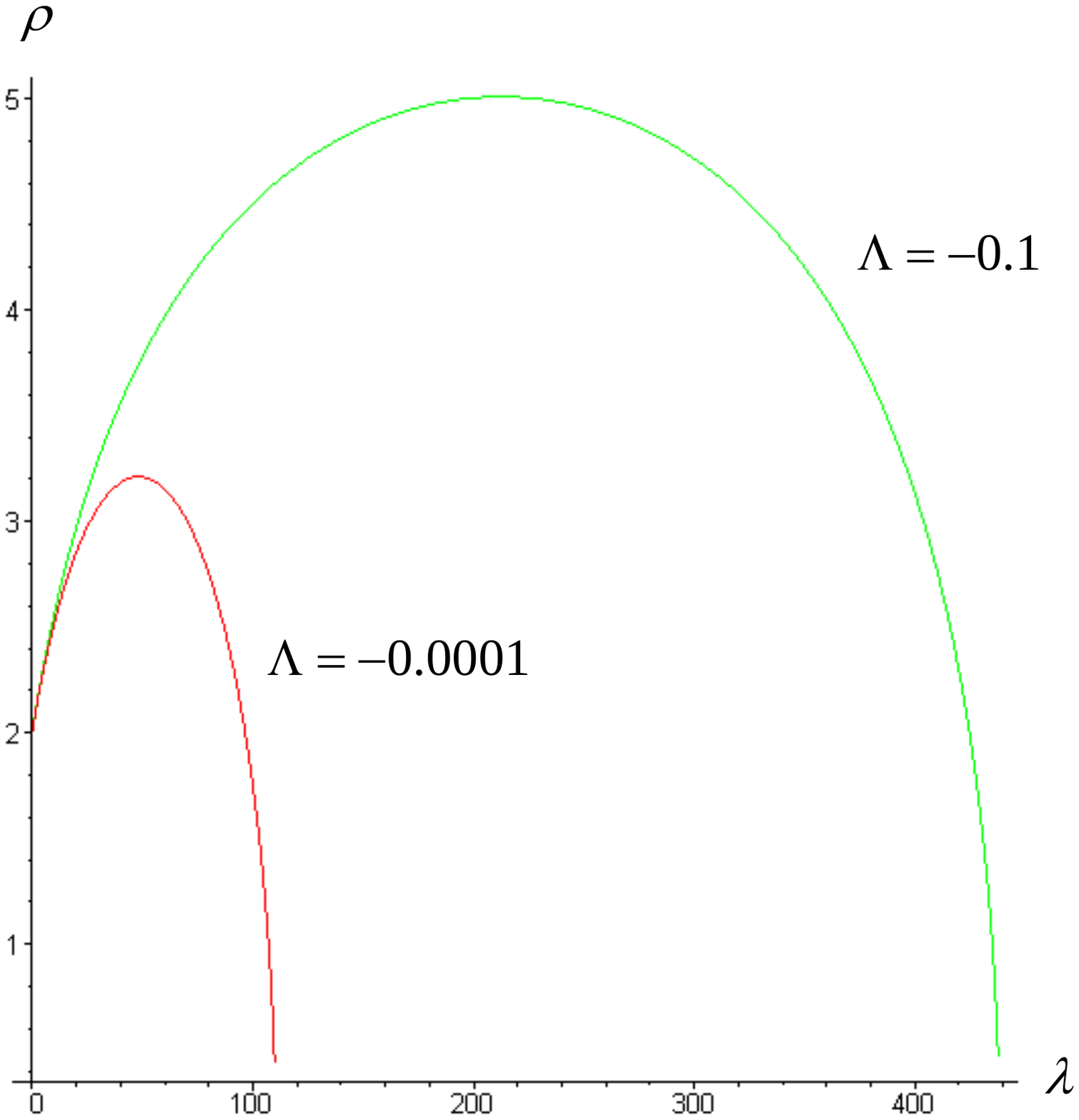} \\
\end{tabular}
\end{minipage}
\begin{minipage}{8cm}
\begin{tabular}{c}
\includegraphics[width=8cm]{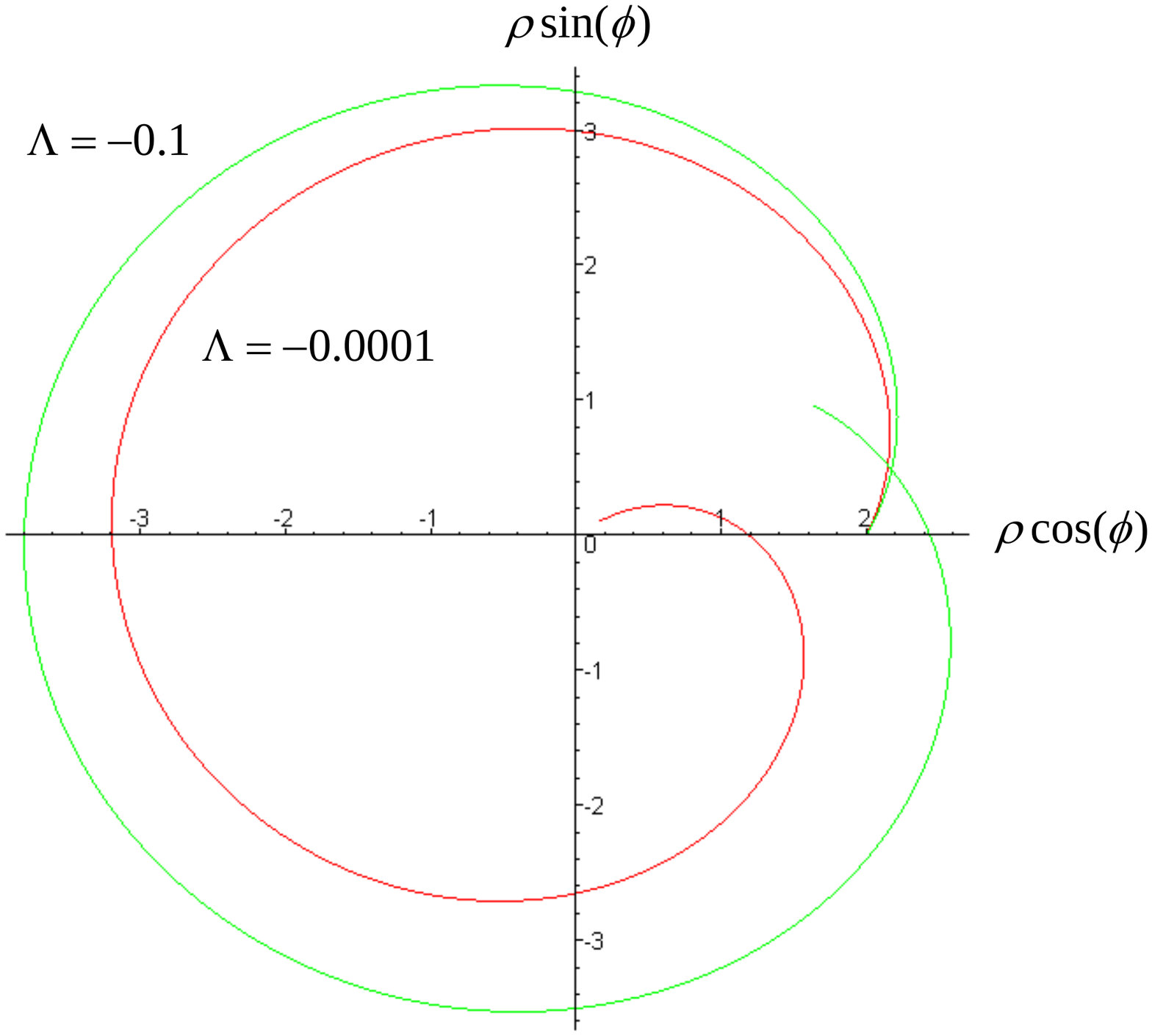}\\
\end{tabular}
\end{minipage}
\caption{ Graphs of the numerical integration of the geodesics' equations along $\rho(\lambda)$, for $E=0.15$, $P_z=\epsilon=0$, $L_z =0.1$, $c=1,$ $\sigma=0.4$, in the cases
$\Lambda=-0.0001$ and $\Lambda=-0.1$, satisfying $E^2< V_{\infty}$. The plots illustrate the fact that $|\Lambda|$ increases the maximum distance of the null geodesics to the axis.}
\label{fig:geod4}
\end{figure}
\end{enumerate}
 \end{enumerate}

\subsubsection{Case $\epsilon\neq 0$}

In this case, (\ref{39}) gives
\begin{equation}
V(\rho)=\epsilon Q^{2/3}P^{-2(1-8\sigma+4\sigma^2)/3\Sigma}+\left(\frac{L_z}{c}\right)^2P^{-2(1-4\sigma)/\Sigma}. \label{52}
\end{equation}
For $\sigma<1/4$, from (\ref{40})-(\ref{43}), we get that the potential $V(\rho)$ always has a minimum at $V^{\star}(\rho_e)=0$ with $V^{\star\star}(\rho_e)>0$ while, for $\sigma\geq 1/4$, there are no equilibrium points, since $V^{\star}(\rho)>0$ and $V(0)=0$, suggesting that we can separate our analysis into two different cases, as follows:

\begin{enumerate}

\item $\sigma<1/4$

In this case, the equation $E^2=V(\rho)$, with $L_z\neq 0$, has two real roots, $\rho_{min}$ and $\rho_{max}$. An incoming timelike particle approaching the axis $z$ is reflected at $\rho=\rho_{min}$, where it attains $\dot\rho=0$, and moves outwards until it attains again $\dot\rho=0$ at $\rho=\rho_{max}$ where it is reflected backwards. This trajectory is repeated endlessly, see examples in Figure \ref{fig:geod5}.

This kind of confinement in the geodesic motion along $\rho$ has been also observed in the van Stockum \cite{Opher}, Lewis \cite{Herrera2} and Lanczos \cite{Pereira} spacetimes.

For $L_z=0$, the incoming radial timelike geodesics hit the $z$ axis, whereas outgoing timelike geodesics reach a maximum finite distance $\rho_{max}$ before turning back to the axis.

\begin{figure}[H]
\begin{minipage}{8cm}
\begin{tabular}{c}
\includegraphics[width=8cm]{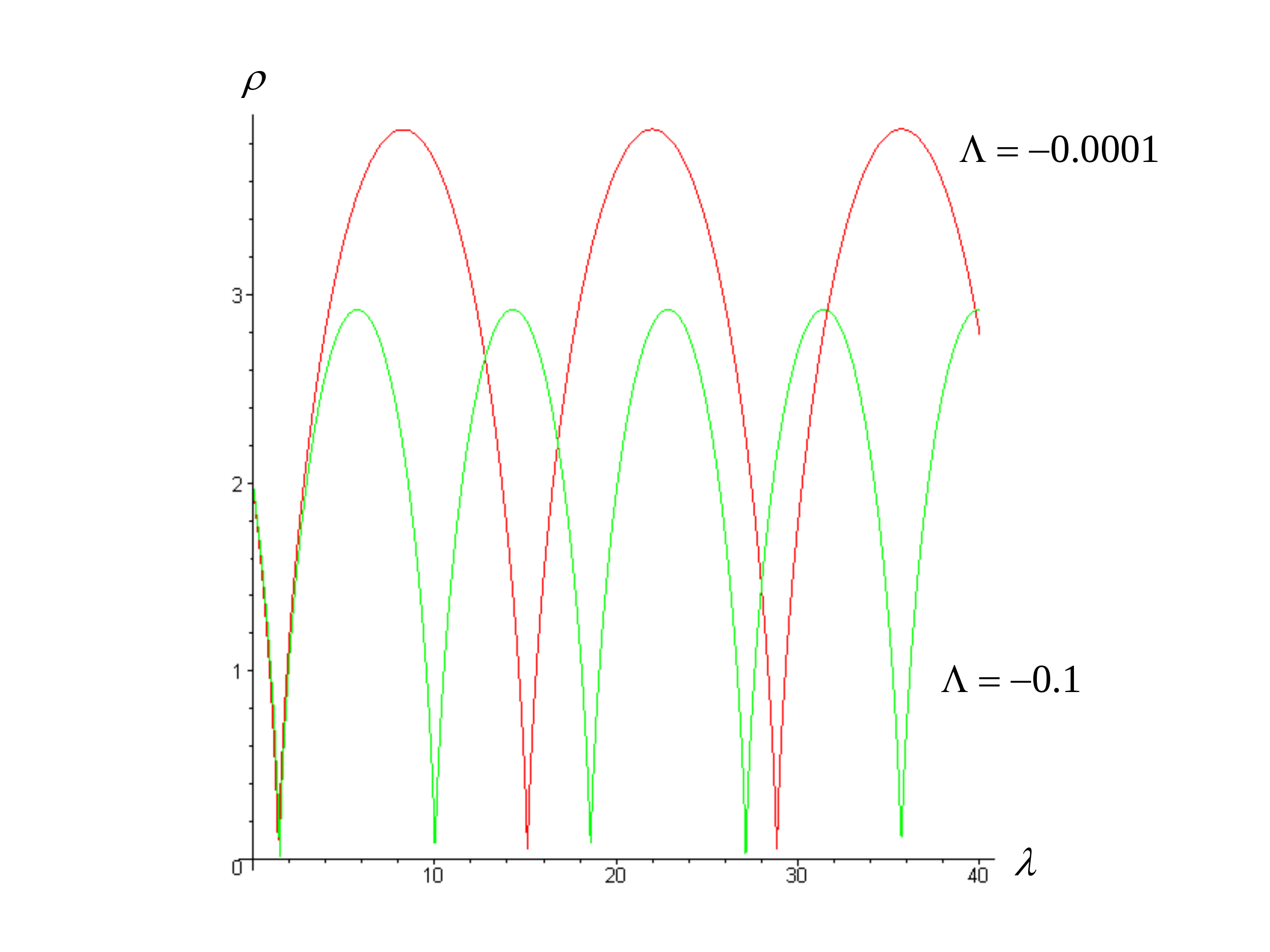} \\
\end{tabular}
\end{minipage}
\begin{minipage}{8cm}
\begin{tabular}{c}
\includegraphics[width=8cm]{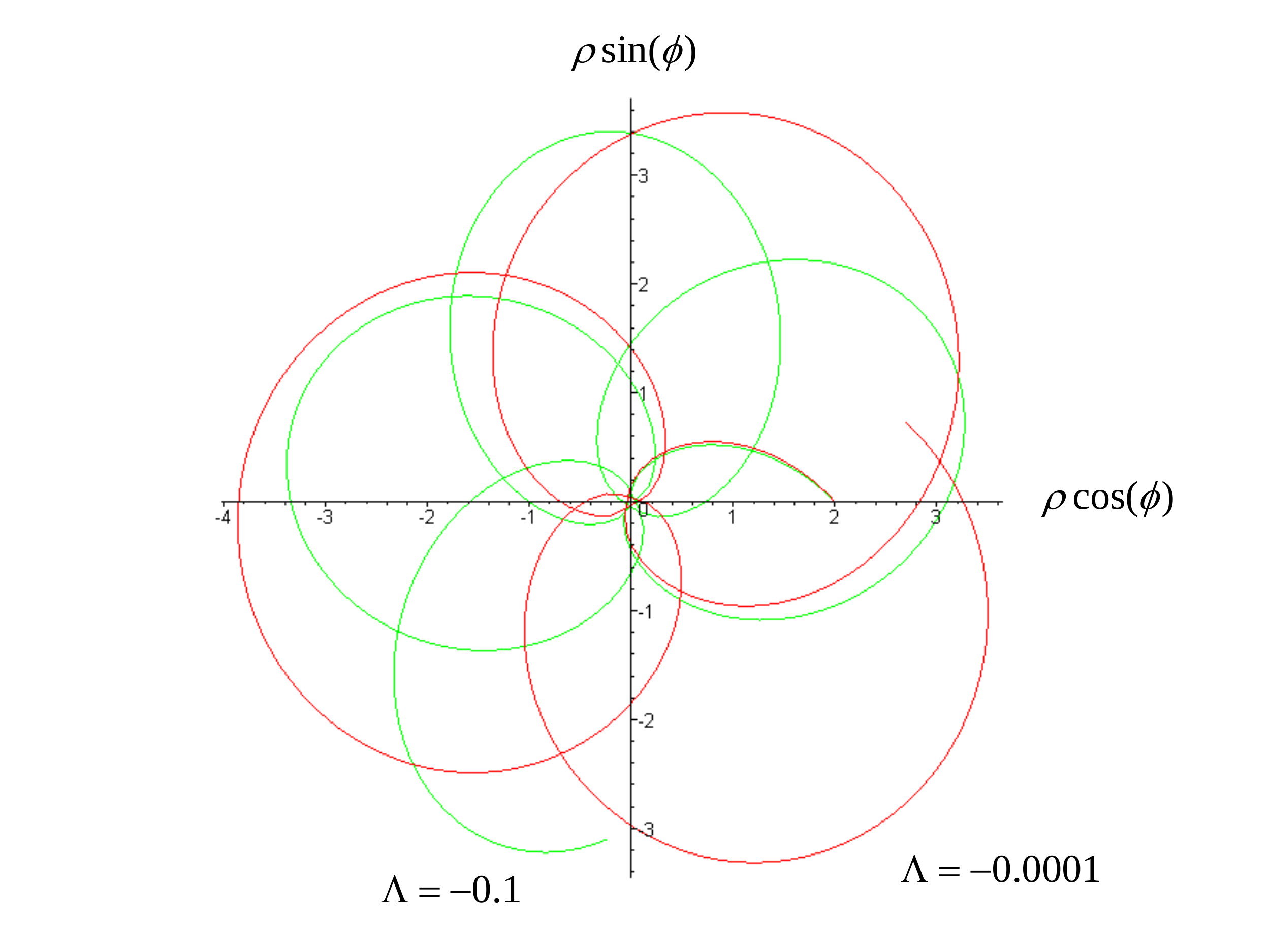}\\
\end{tabular}
\end{minipage}
\caption{ Graphs of the numerical integration of the geodesics' equations along $\rho(\lambda)$, for $E=2$, $P_z=0$, $L_z =\epsilon=c =1$, $\sigma=1/5$, $\lambda\in[0,40],$ in the cases
$\Lambda=-0.0001$ and $\Lambda=-0.1$. In this case, $|\Lambda|$ decreases $\rho_{max}$. This can also be seen, at linear order, by inserting the previous values in (\ref{38B}).}
\label{fig:geod5}
\end{figure}

\item $\sigma\geq 1/4$

For $L_z\neq 0$, as well as for $L_z=0$, incoming radial timelike geodesics hit the $z$ axis, whereas outgoing ones reach a maximum distance $\rho_{max}$ before moving inwards towards the axis. See examples in Figure \ref{fig:geod6}.

\begin{figure}[H]
\begin{minipage}{8cm}
\begin{tabular}{c}
\includegraphics[width=8cm]{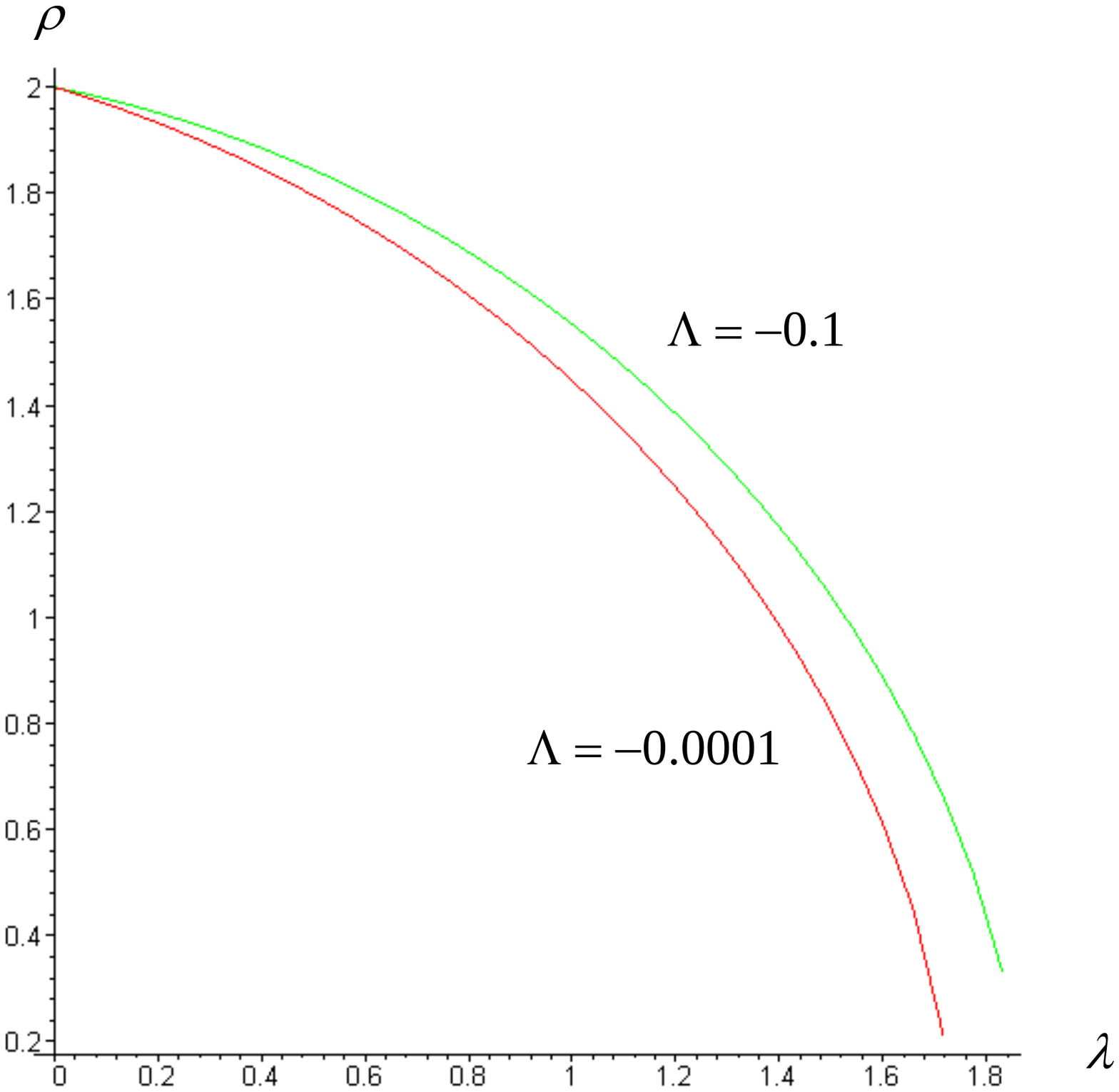} \\
\end{tabular}
\end{minipage}
\begin{minipage}{8cm}
\begin{tabular}{c}
\includegraphics[width=8cm]{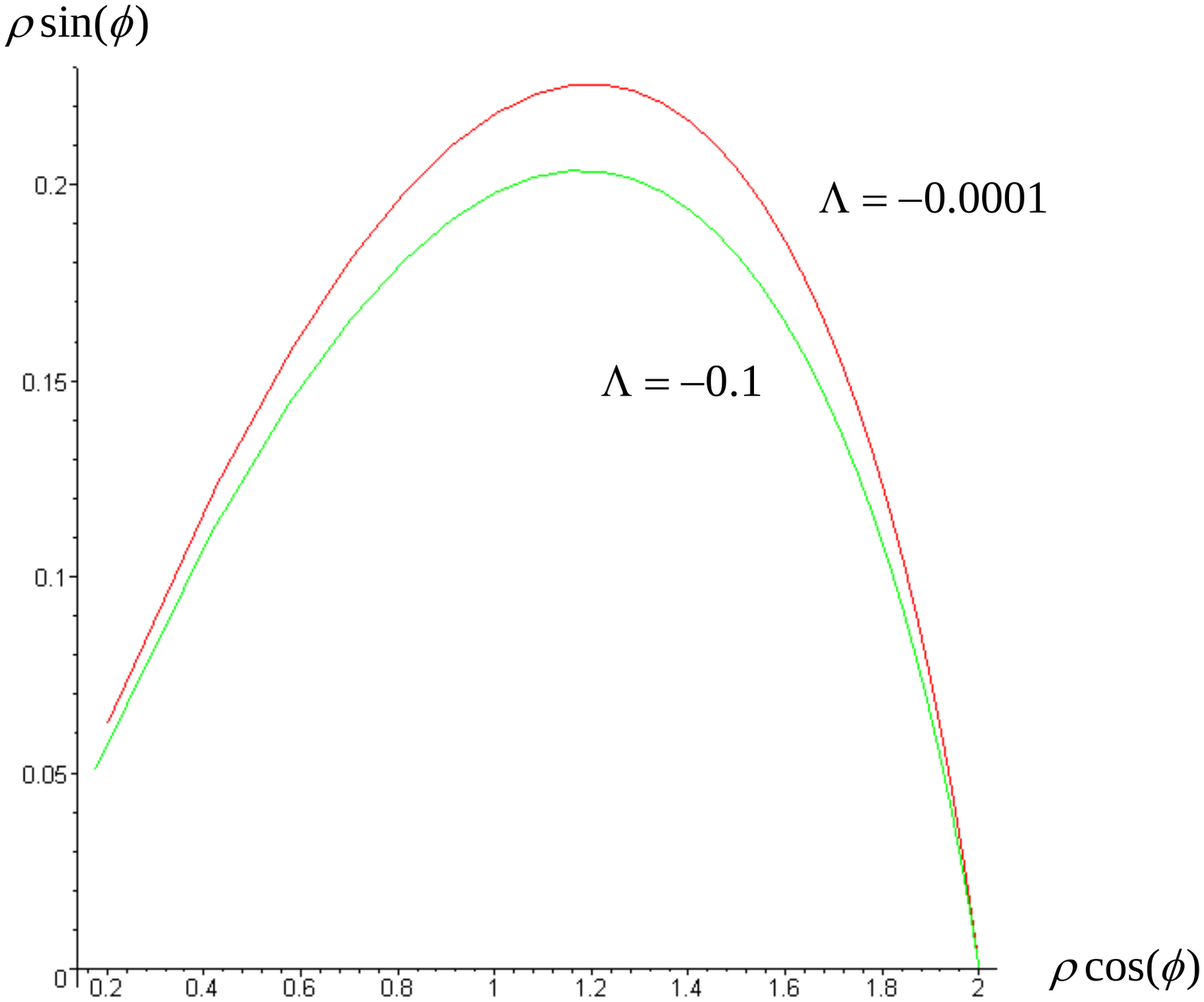}\\
\end{tabular}
\end{minipage}
\caption{ Graphs of the numerical integration of the geodesics' equations along $\rho(\lambda)$, for $E=1.8$, $P_z=0$, $L_z =0.2$, $\epsilon=c =1$, $\sigma=0.4$, in the cases
$\Lambda=-0.0001$ and $\Lambda=-0.1$. }
\label{fig:geod6}
\end{figure}

\end{enumerate}

 Unlike the $\epsilon=0$ case, for $\epsilon\ne 0$, the minimum and maximum distances of the geodesics to the axis can decrease with increasing $|\Lambda|$, depending on the relative magnitudes of $L_z, c$ and $\sigma$. This can be seen by inspecting (\ref{38B}) and in the example plotted in Figure \ref{fig:geod5}.

It is important to note that, as it is demonstrated in \cite{Banerjee0}, for the LC spacetime there is
always geodesic confinement in the case $\epsilon=0, \sigma>1/4$ and $L_z\ne 0$. This can also be confirmed by substituting $P=Q=\rho$ in our equations. We thus conclude that, in that case, for $E^2>V_\infty$, the orbit confinement of null geodesics in the LC spacetime is unstable with respect to the introduction of any $\Lambda<0$.

On the other hand, for $\sigma> 1/4, L_z\ne 0$ and $E^2<V_\infty$, the LC null geodesics' confinement is not broken, and it is therefore stable, against the inclusion of $\Lambda<0$.

\subsection{Non-planar geodesics $(\dot z\ne 0)$}

This is the most general case of geodesics dynamics, which turns out to have, in some sub-cases, a dynamical behaviour along the radial motion which is qualitatively similar to the cases studied in Section 5.1. Nonetheless, we present them in detail because there are some important points to be stressed.

\subsubsection{Case $\epsilon=0$}

From (\ref{39}), we have
\begin{equation}
V(\rho)=P_z^2P^{8\sigma(1-\sigma)/\Sigma}+\left(\frac{L_z}{c}\right)^2P^{-2(1-4\sigma)/\Sigma}, \label{35}
\end{equation}
and,
asymptotically, for $\rho\rightarrow\infty$, we have from (\ref{35})
\begin{equation}
V_{\infty}=P_z^2\left(\frac{2}{\sqrt{3|\Lambda|}}\right)^{8\sigma(1-\sigma)/\Sigma}
+\left(\frac{L_z}{c}\right)^2\left(\frac{2}{\sqrt{3|\Lambda|}}\right)^{-2(1-4\sigma)/\Sigma}. \label{57}
\end{equation}
While in the planar case the analysis was splitted into three cases, since the case $\sigma=1/4$ was treated separately, here we consider the two different cases:

\begin{enumerate}
\item $\sigma<1/4$
\begin{enumerate}
\item If $E^2>V_{\infty}$, a null particle approaches the axis with decreasing negative acceleration, $\ddot\rho<0$, and increasing speed, see (\ref{38}) and (\ref{44}). The particle attains its maximum speed at $\ddot\rho=0$ and from there onwards diminishes its speed, since $\ddot\rho>0$, until it arrives at its minimum distance from the axis at $E^2=V(\rho)$, where it has vanishing speed. From there on, the null particle is reflected escaping to infinity.
If $L_z=0$, incoming null particles hit the $z$ axis. Examples are plotted in Figure \ref{fig:geod7}.

\begin{figure}[H]
\begin{minipage}{8cm}
\begin{tabular}{c}
\includegraphics[width=8cm]{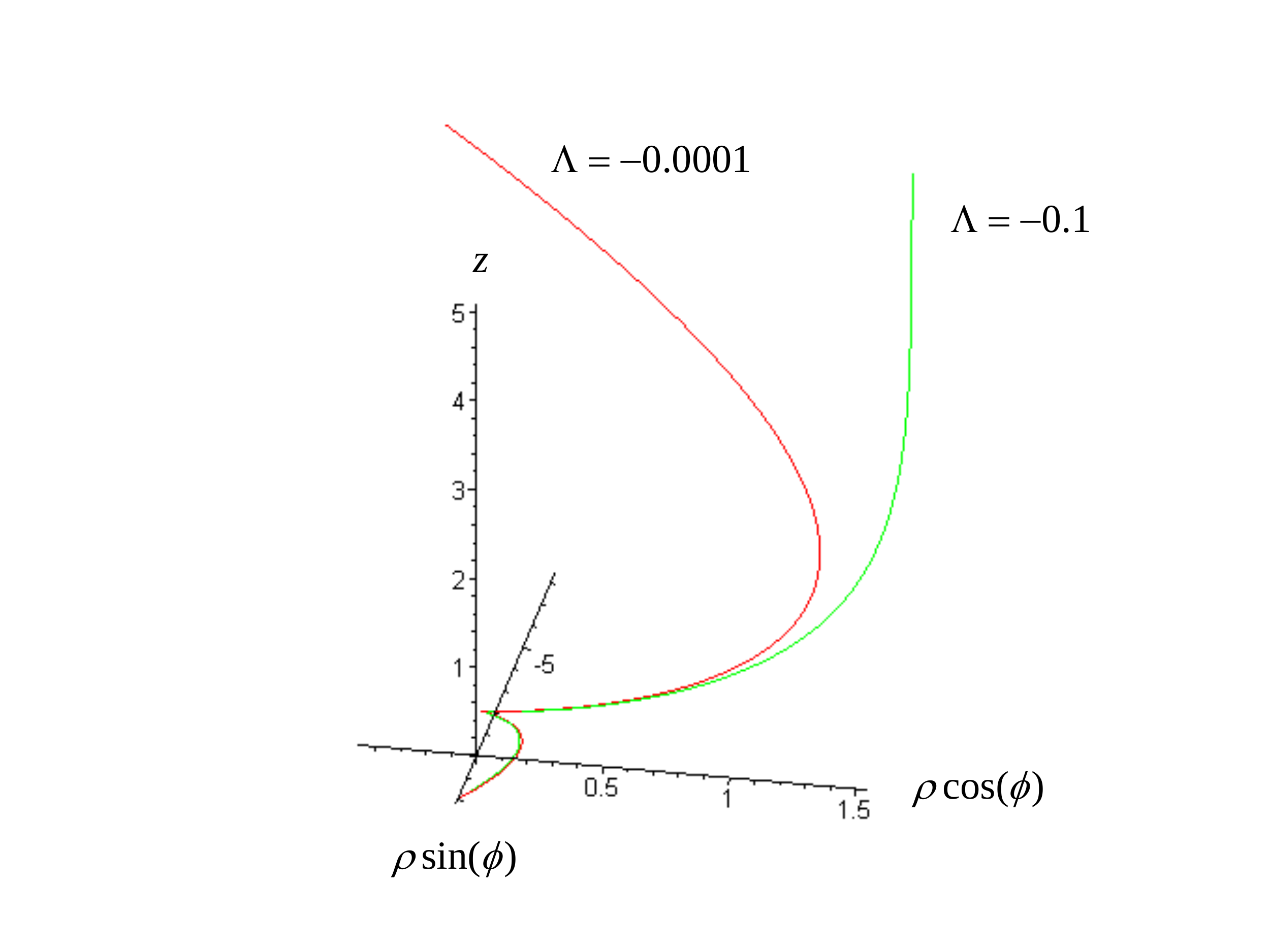} \\
\end{tabular}
\end{minipage}
\begin{minipage}{8cm}
\begin{tabular}{c}
\includegraphics[width=8cm]{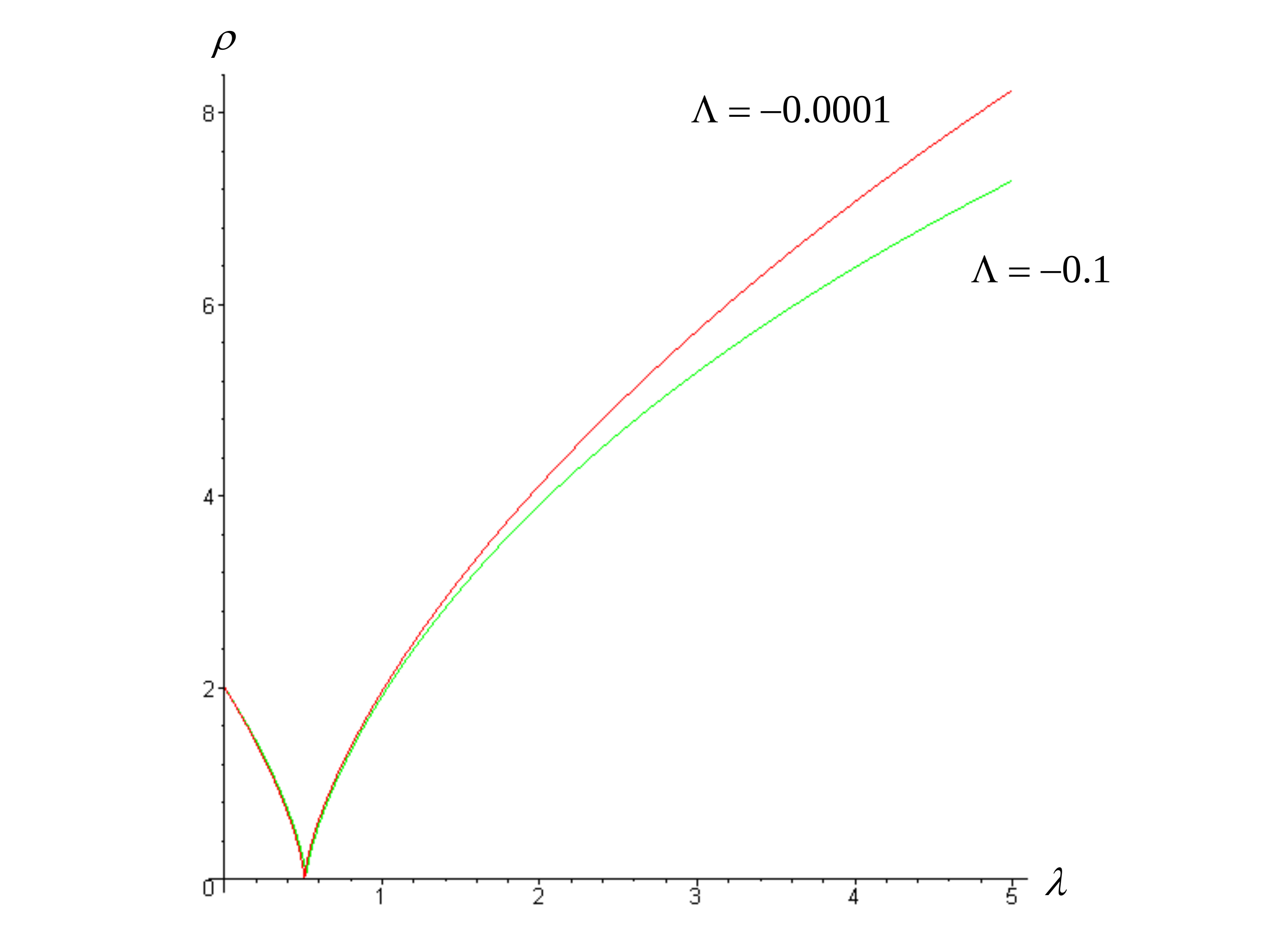}\\
\end{tabular}
\end{minipage}
\caption{ Graphs of the numerical integration of the non-planar null geodesics' equations along $\rho(\lambda)$, for $E=4$, $L_z =P_z=c=1$, $\epsilon =0$, $\sigma=1/5$, $\lambda\in[0,5],$ for $\Lambda=-0.0001$ and for $\Lambda=-0.1$. In this example $E^2>V_{\infty}$. }
\label{fig:geod7}
\end{figure}

\item If $E^2=V_{\infty}$, a null particle has a similar radial motion as in the previous case, but with the difference that its energy $E$ is the minimum required for the particle to reach an infinite distance from the axis.

\item If $E^2<V_{\infty}$, the null particle attains zero speed for the two roots of $E^2=V(\rho)$, $\rho_{min}$ and $\rho_{max}$. The particle is reflected from $\rho_{min}$ to $\rho_{max}$ where it is reflected backwards to $\rho_{min}$. This motion is repeated endlessly, which characterises a confinement of the particle along $\rho$. Numerical examples are shown in Figure \ref{fig:geod8}.

It is interesting to compare this confinement to the one produced for $P_z=0$ and $\epsilon\neq 0$ in Section 5.1.2. In this case, with $P_z\neq 0$ and $\epsilon=0$, one might interpret the motion of the null particle along $\rho$ as becoming endowed with a kind of "inertial mass" produced by its momentum along $z$, $P_z$.

If $L_z=0$, incoming null particles hit the $z$ axis, whereas outgoing particles reach a maximal distance, $\rho_{max}$, where $\dot\rho=0$. In this case, it is easy to get from (\ref{38B})
\begin{equation}
\rho_{max}\approx\rho_{LCmax} +\;\frac{|\Lambda|}{4}\rho_{LCmax}^{(3-6\sigma+12\sigma^2)/\Sigma} ,
\end{equation}
which shows, at linear order, how increasing $|\Lambda|$ increases $\rho_{max}$.

\begin{figure}[H]
\begin{minipage}{8cm}
\begin{tabular}{c}
\includegraphics[width=8cm]{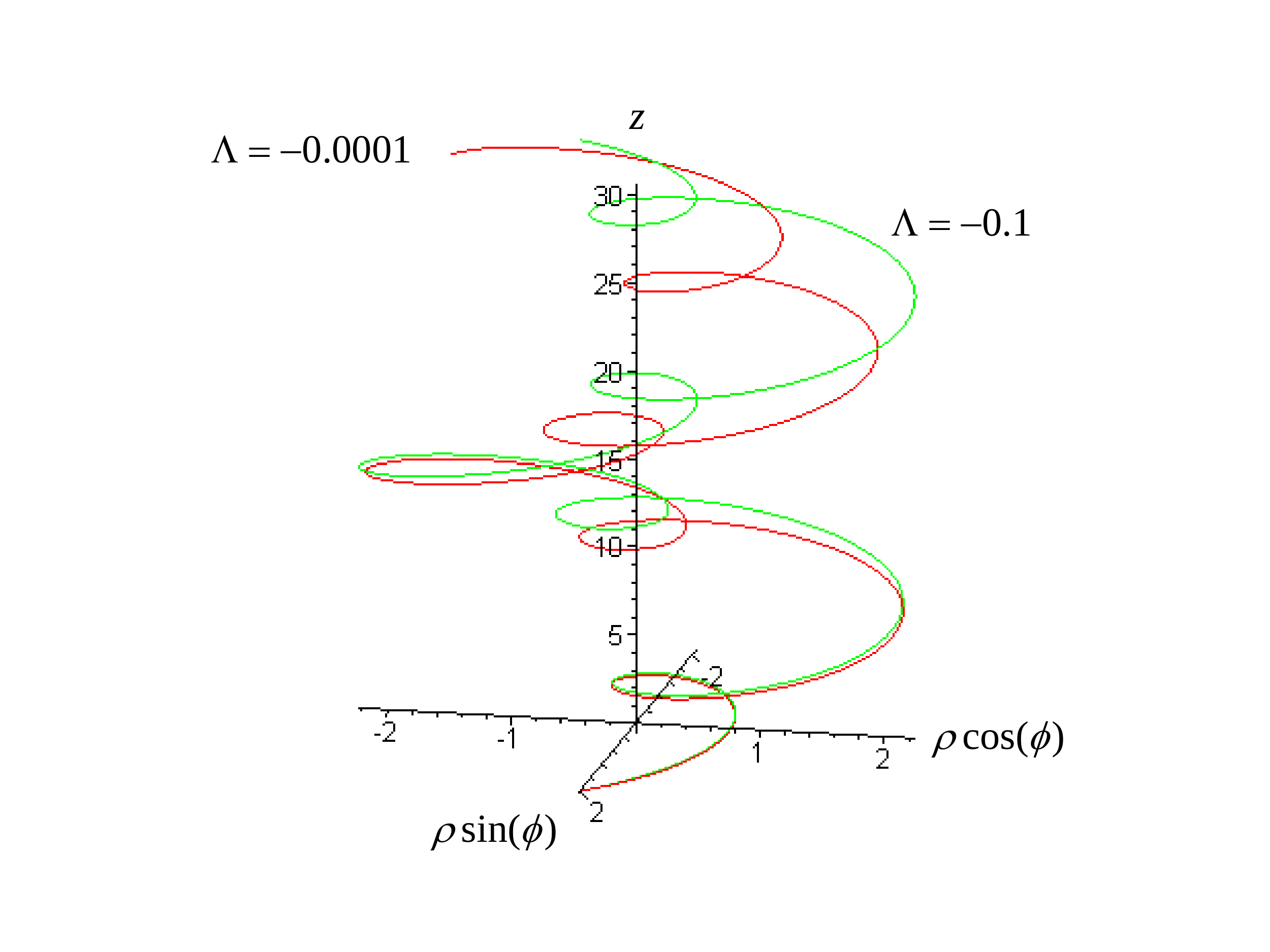} \\
\end{tabular}
\end{minipage}
\begin{minipage}{8cm}
\begin{tabular}{c}
\includegraphics[width=8cm]{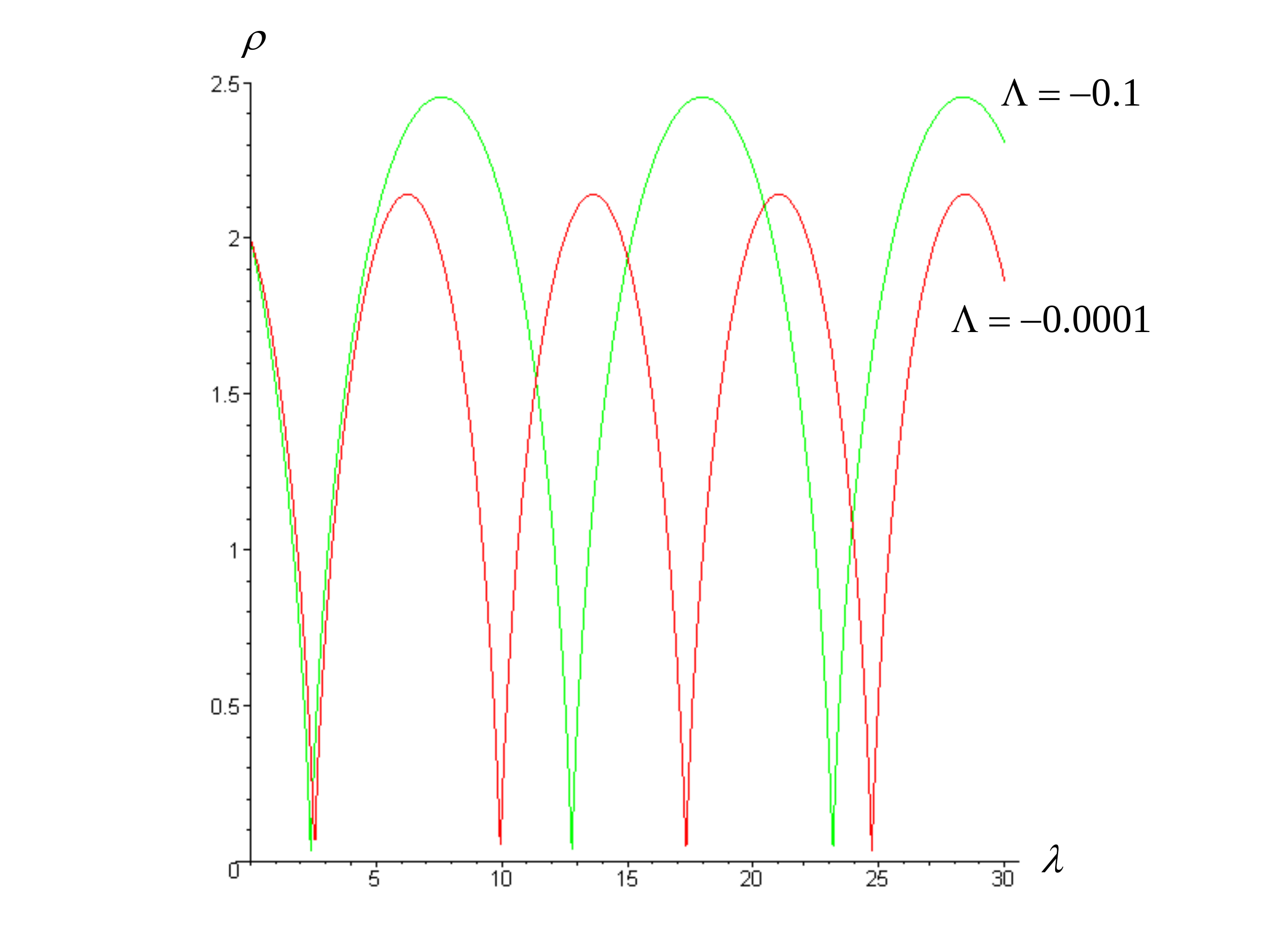}\\
\end{tabular}
\end{minipage}
\caption{Graphs of the numerical integration of the non-planar null geodesics' equations along $\rho(\lambda)$, for $E=1.7$, $L_z =P_z=c=1$, $\epsilon =0$, $\sigma=1/5$, $\lambda\in[0,30],$ for $\Lambda=-0.0001$ and for $\Lambda=-0.1$. In this example $E^2<V_{\infty}$. The graphs on the right represent the evolution of the radii of geodesics confined between $\rho_{min}$ and $\rho_{max}$. }
\label{fig:geod8}
\end{figure}

\end{enumerate}

\item $\sigma\geq 1/4$
\begin{enumerate}
\item If $E^2>V_{\infty}$, incoming null particles hit the $z$ axis, whereas outgoing ones escape to infinity, see an example in Figure \ref{fig:geod9}.

\begin{figure}[H]
\begin{minipage}{8cm}
\begin{tabular}{c}
\includegraphics[width=8cm]{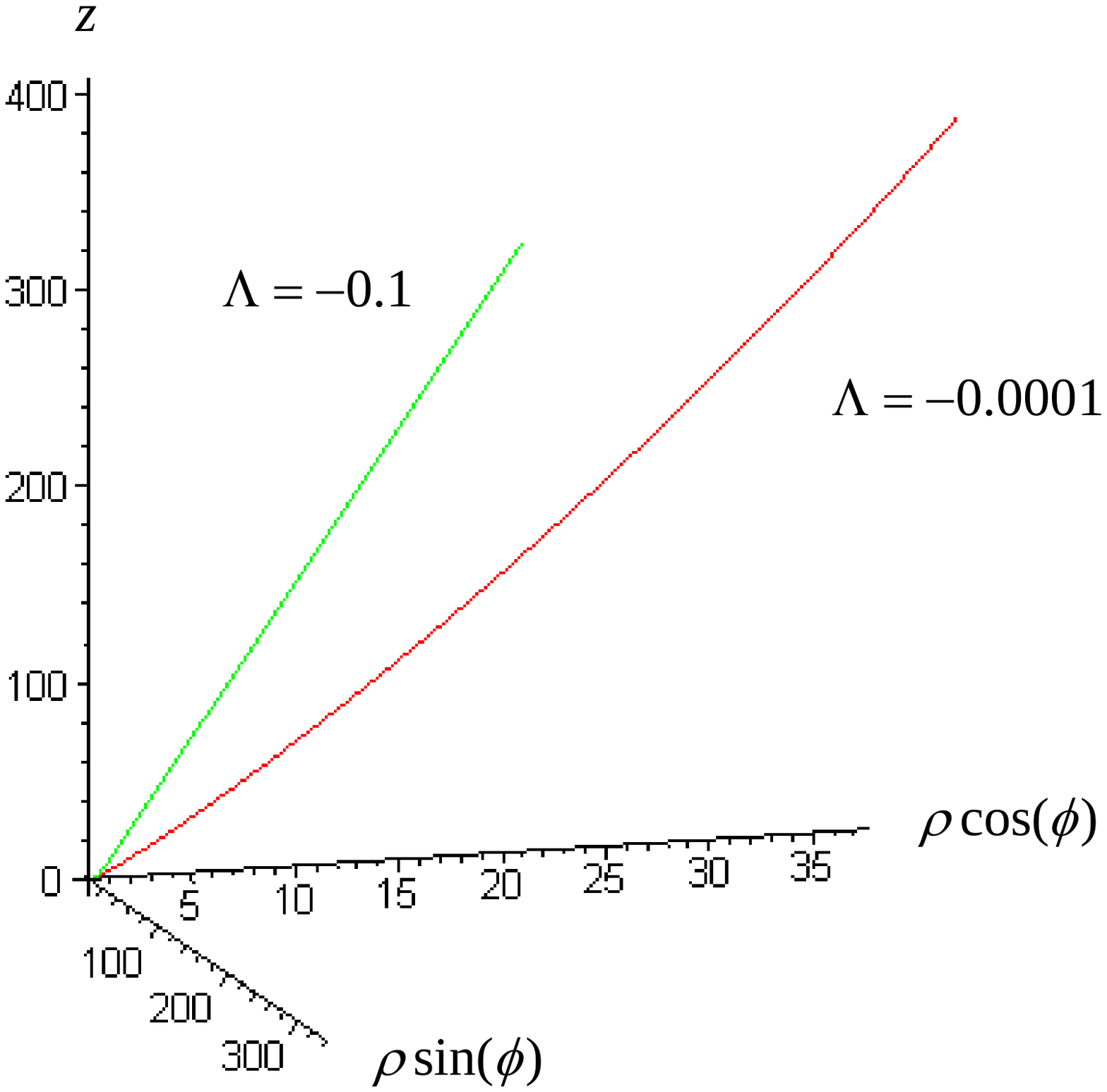} \\
\end{tabular}
\end{minipage}
\begin{minipage}{8cm}
\begin{tabular}{c}
\includegraphics[width=8cm]{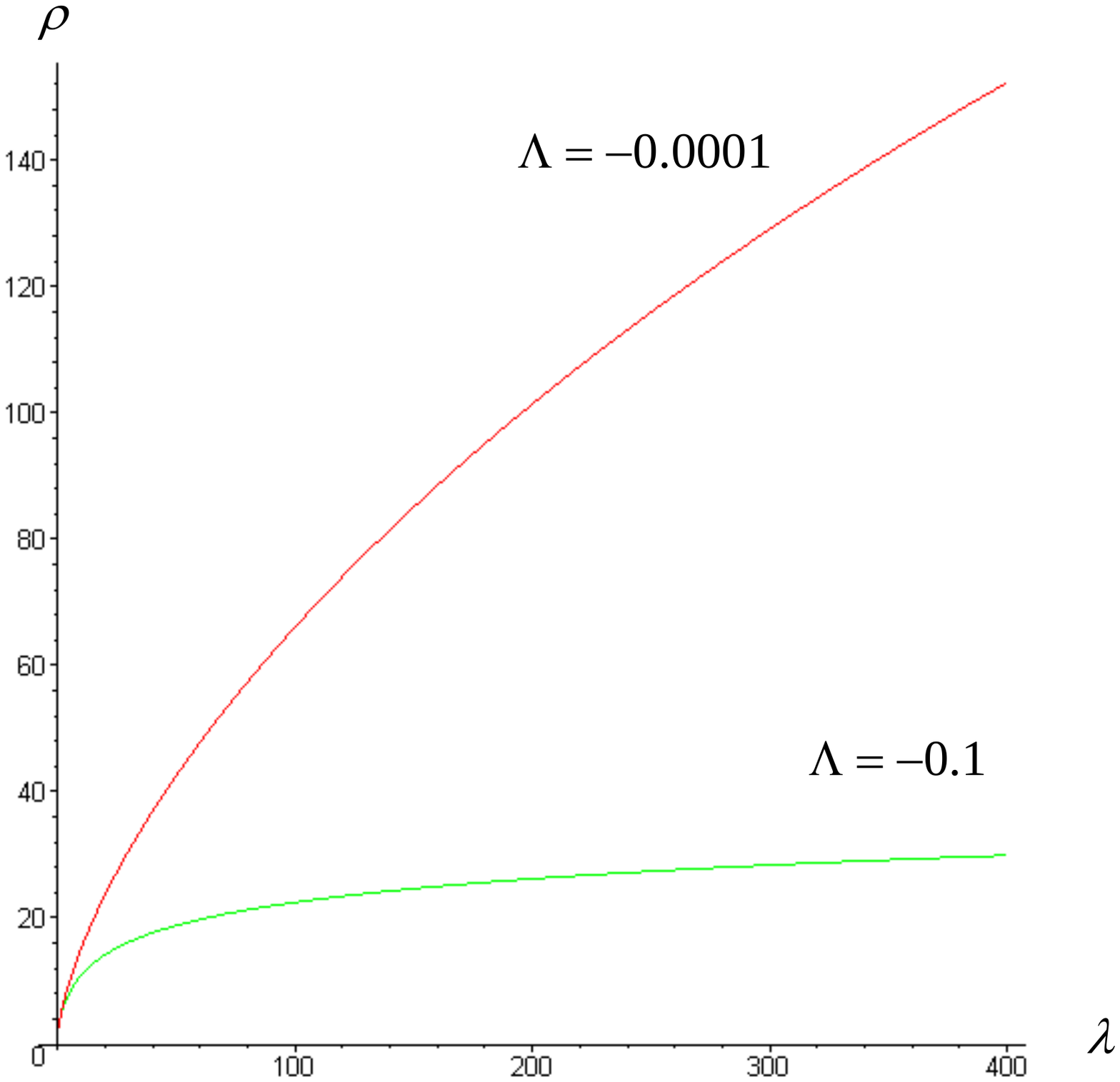}\\
\end{tabular}
\end{minipage}
\caption{Graphs of the numerical integration of the geodesics' equations along $\rho(\lambda)$, for $E=4$, $P_z=0.1$, $L_z =0.05$, $\epsilon =0$, $c =1$, $\sigma=0.25$, in the cases
$\Lambda=-0.0001$ and $\Lambda=-0.1$. In those cases $E^2> V_\infty$.}
\label{fig:geod9}
\end{figure}

\item If $E^2=V_{\infty}$, incoming null particles also hit the $z$ axis, whereas outgoing ones have the minimum energy to reach infinity.

\item If $E^2<V_{\infty}$, incoming null particles hit the axis $z$ and outgoing ones reach a maximum distance $\rho_{max}$, where $\dot\rho=0$. See a numerical example in Figure \ref{fig:geod10}.  By inspecting  (\ref{38B}) it is easy to quantify, at linear order, the increase in $\rho_{max}$ with increasing $|\Lambda|$, in this case.

\begin{figure}[H]
\begin{minipage}{8cm}
\begin{tabular}{c}
\includegraphics[width=8cm]{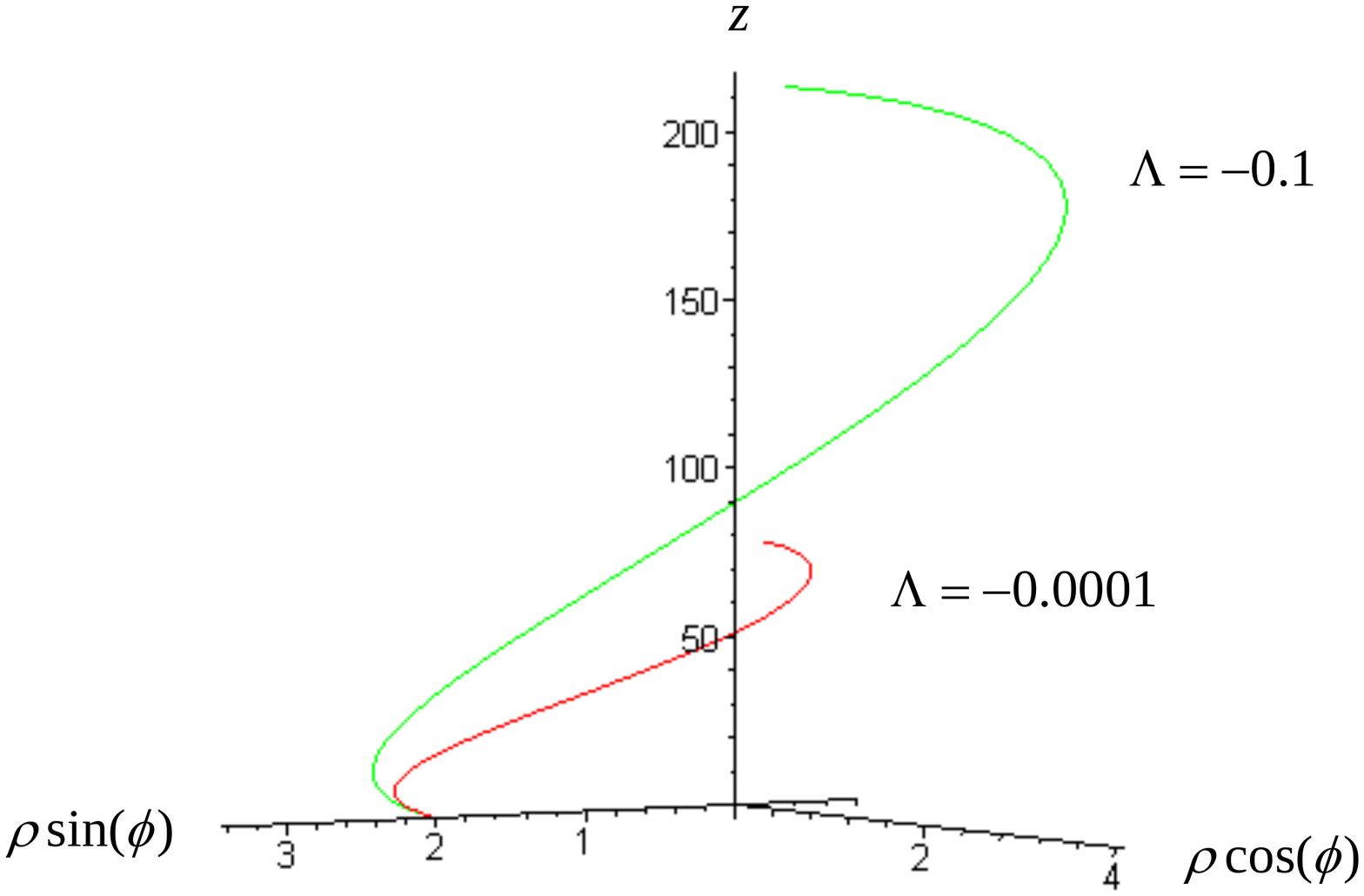} \\
\end{tabular}
\end{minipage}
\begin{minipage}{8cm}
\begin{tabular}{c}
\includegraphics[width=8cm]{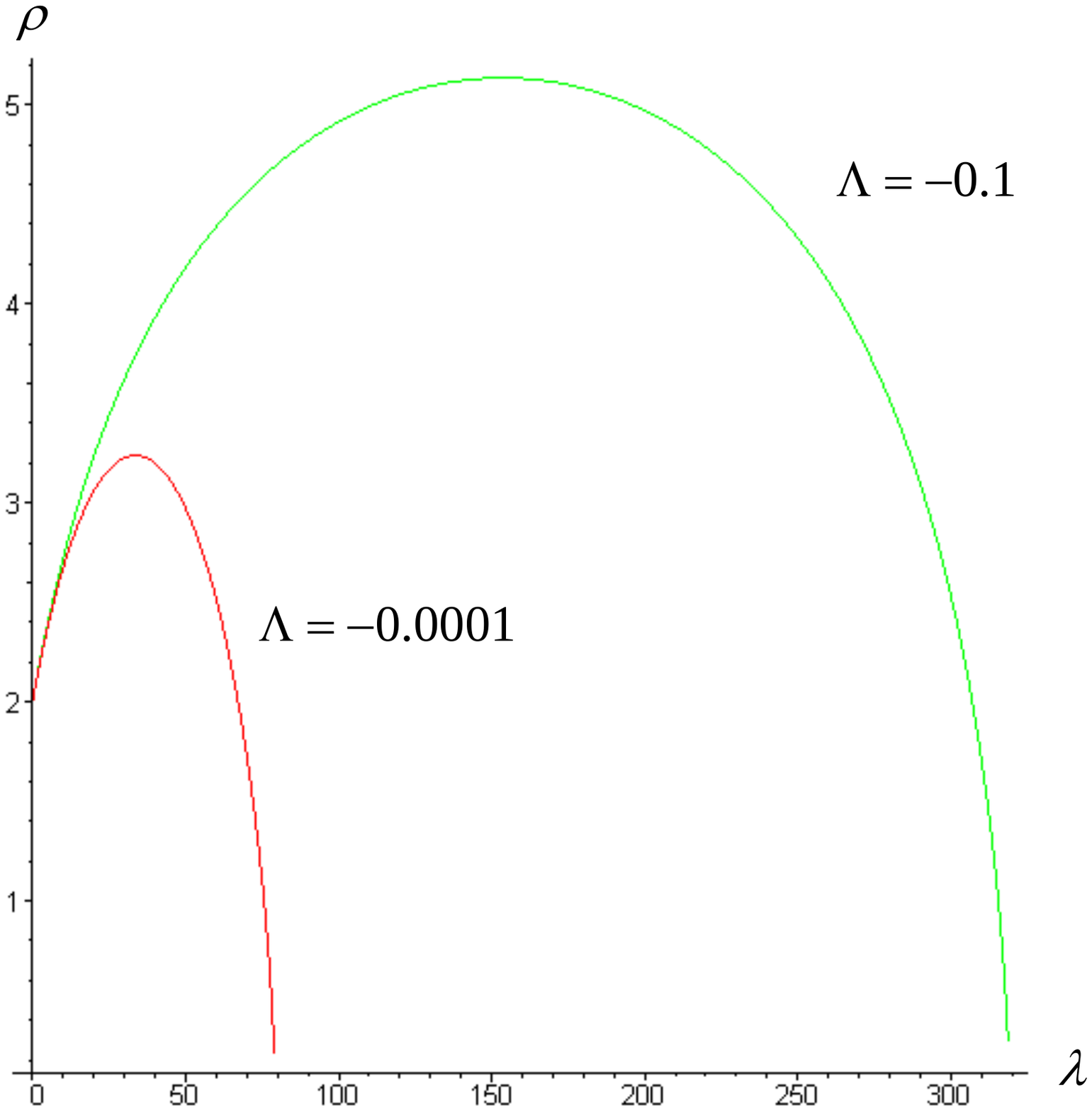}\\
\end{tabular}
\end{minipage}
\caption{ Graphs of the numerical integration of the geodesics' equations along $\rho(\lambda)$, for $E=0.2$, $P_z=0.1$, $L_z =0.05$, $\epsilon =0$, $c =1$, $\sigma=0.25$, in the cases
$\Lambda=-0.0001$ and $\Lambda=-0.1$. In those cases $E^2< V_\infty$.}
\label{fig:geod10}
\end{figure}

\end{enumerate}
\end{enumerate}

 An interesting difference between  planar and non-planar null geodesics is that, while in the former case $|\Lambda|$ always increases the extreme distances of the geodesics to the axis and tends to destabilise their dynamics, in the latter case the effect of $|\Lambda|$ on the geodesics's orbits depends on the relative magnitudes of $P_z, L_z, c$ and $\sigma$. This can be seen through (\ref{38B}), as linear order effect, and in the non-linear examples of Figures \ref{fig:geod7}, \ref{fig:geod8}, \ref{fig:geod9} and \ref{fig:geod10}.

Another important point is that, in the LC metric with $\epsilon=0$ and $P_z\neq 0$, after substituting $P=Q=\rho$ in our formulae, we always get geodesic confinement in the radial direction (see also \cite{Herrera2}), while in LT this is not so. Indeed, in the LT metric, for any $\sigma$, as long as $E^2\ge V_\infty$, the null geodesics escape to infinity and we therefore conclude that, in those cases, the geodesic motion in the LC metric is unstable with respect to the introduction of any values of $\Lambda<0$.

On the other hand, if $E^2< V_\infty$, we find that the LC geodesics' confinement is maintained after including $\Lambda<0$.

\subsubsection{Case $\epsilon\neq 0$}

\begin{figure}[H]
\begin{minipage}{8cm}
\begin{tabular}{c}
\includegraphics[width=8cm]{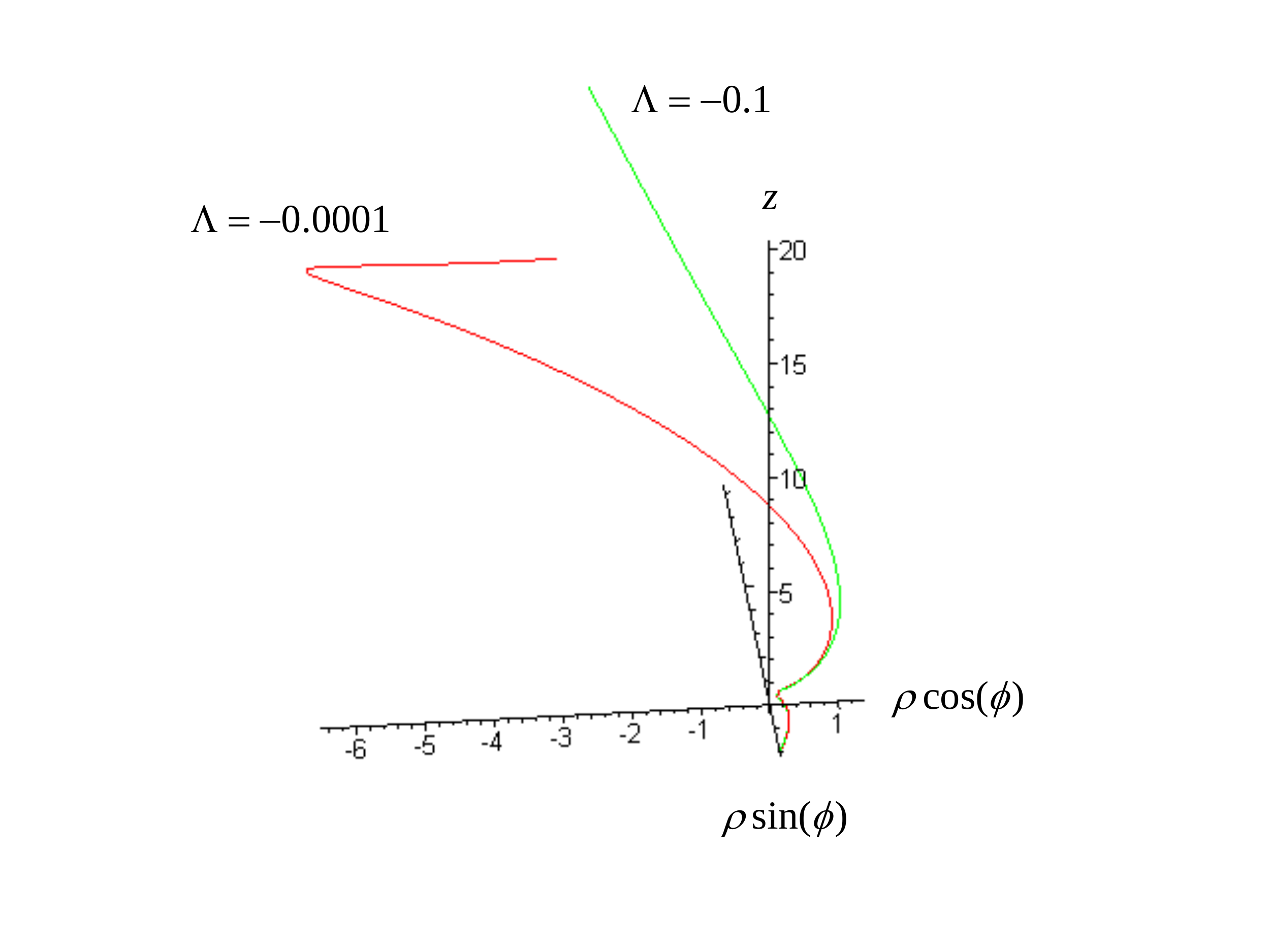} \\
\end{tabular}
\end{minipage}
\begin{minipage}{8cm}
\begin{tabular}{c}
\includegraphics[width=8cm]{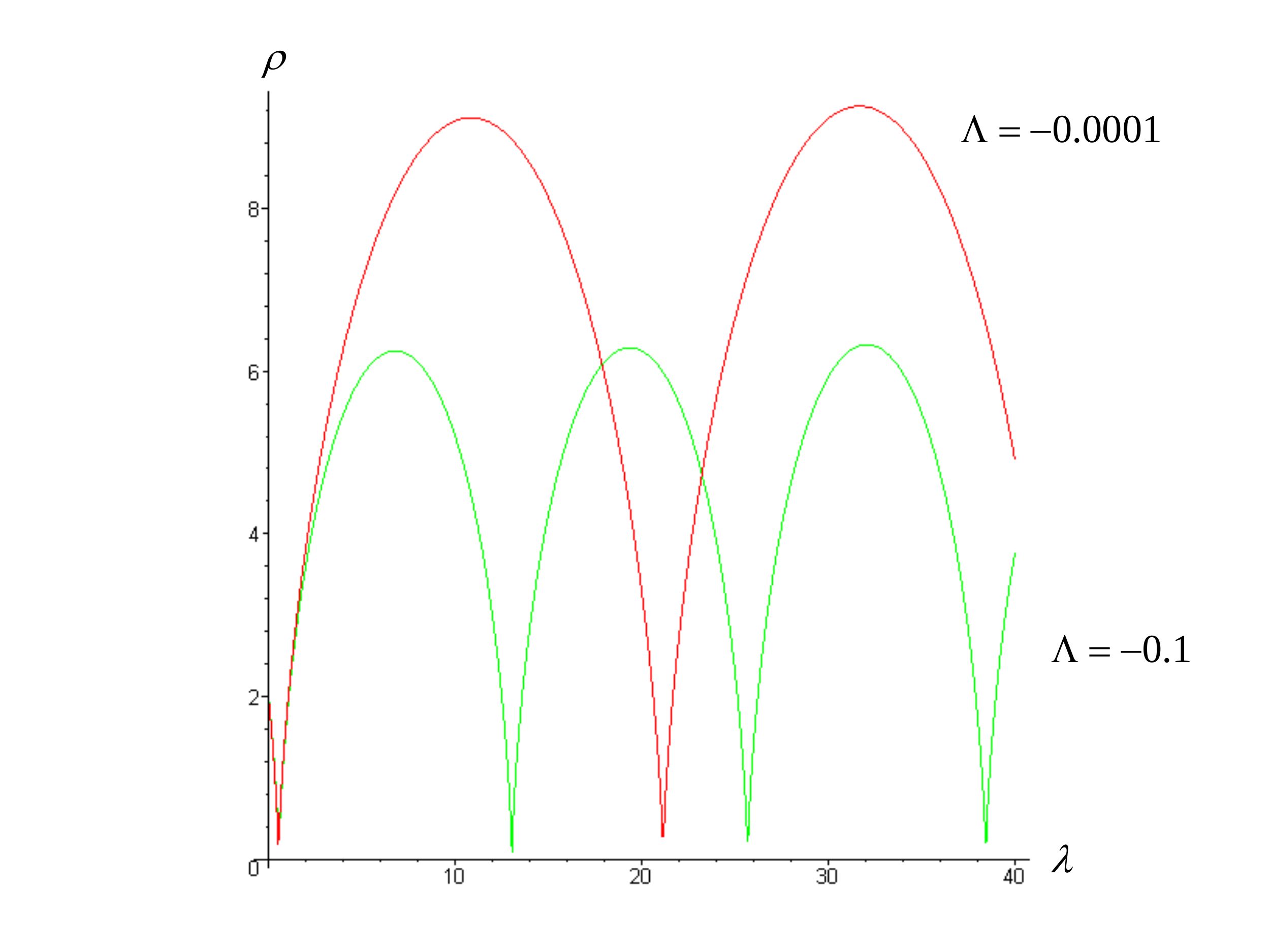}\\
\end{tabular}
\end{minipage}
\caption{ Graphs of the numerical integration of the non-planar geodesics' equations along $\rho(\lambda)$, for $E=4$, $L_z =P_z=c=1$, $\epsilon =1$, $\sigma=1/5$, $\lambda\in[0,40],$ in the cases
$\Lambda=-0.0001$ and $\Lambda=-0.1$. In this example, increasing $|\Lambda|$ decreases $\rho_m$.}
\label{fig:geod11}
\end{figure}

In this case, using formulae (\ref{39})-(\ref{44}), one can show that there is always geodesic confinement
in the radial direction of the particle, like in the LC spacetime \cite{Herrera2}.  Examples are plotted in Figure \ref{fig:geod11}.
Interestingly, the geodesic confinement along $\rho$ for all values of $\sigma$ or $L_z$ (including $L_z=0$) is a distinguishing feature from all the previous $\epsilon=0$ cases. 

The values of $\rho_{min}$ and $\rho_{max}$ are given by the zeros of (\ref{38}). At linear order in $|\Lambda|$, the $\rho_m$ are related to the extreme values of $\rho$ of the LC metric, $\rho_{LCm}$, through (\ref{38B}).
In general, the influence of $|\Lambda|$ on the values of  $\rho_{min}$ and $\rho_{max}$ depends on the relative values of the constants $P_z, L_z, c$ and $\sigma$, as in some of the former cases. By fixing the value of some of the constants, though, one can extract useful information independently from the values of the remaining constants.
For example, for $\epsilon=1$, in the limit cases $\sigma=1/2$ and $\sigma=0$ (with $L_z\ne 0$ and any $P_z$), respectively, (\ref{38B}) gives
\begin{equation}
\rho_{m}\approx\rho_{LCm} +\;\frac{|\Lambda|\rho^3_{LCm}}{4}\left[\left(\frac{L_z}{c}\right)^2+P_{z}^2\right]  \left[\left(\frac{L_z}{c}\right)^2+1+P_{z}^{2}\right]^{-1},
\end{equation}
and
\begin{equation}
\rho_{m}\approx\rho_{LCm} +\;\frac{|\Lambda|\rho^3_{LCm}}{4}\left[1+\left(\frac{c\rho_{LCm}}{L_z}\right)^{2}\right],
\end{equation}
which reveals, in those cases, at linear order, how increasing values of $|\Lambda|$ increase the extreme values of the geodesics' distance to the axis.

However, as in previous cases, here increasing $|\Lambda|$ can decrease $\rho_m$ for some particular values of the constants, as illustrated in the example of Figure \ref{fig:geod11}.

\section{Conclusion}

 In this paper, we have investigated the dynamics of geodesics in cylindrically symmetric vacuum LT metrics with $\Lambda<0$.
 In particular we have addressed the question of the stability of the geodesics' motion against the introduction of arbitrarily small values of $\Lambda<0$.

We have found that, for planar null geodesics, increasing $|\Lambda|$
tends to increase the minimum and maximum radial distances to the axis of the confined geodesics. Non-null geodesics are always confined and the effect of $|\Lambda|$ on their extreme distances to the axis depends, in general, on the relative magnitudes of $P_z, L_z, c$ and $\sigma$.  In order to quantify that effect in those cases, we have used linear perturbations in $\Lambda$.

In turn, for some non-planar null geodesics with arbitrary $0\le \sigma\le 1/2$ and some planar null geodesics with $\sigma> 1/4$, the inclusion of any $\Lambda<0$ breaks the orbit confinement of the $\Lambda=0$ geodesics. In this sense, those null geodesics are unstable against the introduction of an arbitrarily small $\Lambda<0$.

A key ingredient in our investigation was the use of an appropriate potential function which enabled a qualitative analysis of the geodesics' system of equations.
To illustrate our findings, we did numerical simulations of the full system of geodesics' equations and plotted some examples which we compared with our stability results.

 Finally, we recall that although families of spatially confined planar null geodesics were known to exist, we have found no trapped cylinders in the LT spacetime. An interesting side result of this paper is the clarification of this issue, as we have shown that the planar null geodesics which are confined are {\it non-radial}, while all outgoing {\it radial} planar null geodesics escape to infinity.


Here, we did not consider the case $\Lambda>0$. In that case, the LT metric contains a second curvature singularity and represents, at most,  the gravitational field in the region between two cylindrical sources.  However, although partial results have been obtained in \cite{GP, Brito}, it is still an open problem to find a metric which can represent, simultaneously, the two sources of the $\Lambda>0$ LT spacetime.

\section*{Acknowledgments}
IB and FM thank CMAT, Univ. Minho, for support through the FEDER Funds-COMPETE and FCT Project Est-C/MAT/UI0013/2011. FM is also supported by FCT projects\\ PTDC/MAT/108921/2008 and CERN/FP/123609/2011 and thanks the warm hospitality from Instituto de F\'isica, UERJ, Rio de Janeiro, where this work was completed. MFAdaSilva acknowledges the financial support from FAPERJ (no. E-26/171.754/2000, E-26/171.533.2002, E-26/170.951/2006, E-26/110.432/2009 and E-26/111.714/2010), Conselho Nacional de Desenvolvimento Cient\'{i}fico e Tecnol\'ogico - CNPq - Brazil (no. 450572/2009-9, 301973/2009-1 and 477268/2010-2) and Financiadora de Estudos e Projetos - FINEP - Brazil.

\end{document}